\documentclass[article,longauth]{aa} 

\usepackage{graphicx}
\usepackage{lineno}

\usepackage{txfonts}
\usepackage[colorlinks=true,linkcolor=blue,citecolor=blue]{hyperref} 
\usepackage[usenames,dvipsnames]{xcolor}
\usepackage{float}
\usepackage{natbib}
\bibpunct{(}{)}{;}{a}{}{,}
\usepackage{soul}
\usepackage{orcidlink}
\usepackage{arydshln}
\usepackage[utf8]{inputenc}
\usepackage{pifont} 
\newcommand{\cmark}{\ding{51}} 

\begin{document} 
\nolinenumbers

\title{JWST-TST High Contrast: Medium-resolution spectroscopy reveals a carbon-rich circumplanetary disk around the young accreting exoplanet Delorme 1\,AB\,b}

\subtitle{}
\author{Mathilde Mâlin\orcidlink{0000-0002-2918-8479}\inst{\ref{jhu},\ref{stsci}}, 
    Kimberly Ward-Duong\orcidlink{0000-0002-4479-8291}\inst{\ref{smith}}, 
    Sierra L. Grant\orcidlink{0000-0002-4022-4899}\inst{\ref{carnegie}}, 
    Nicole Arulanantham\orcidlink{0000-0003-2631-5265}\inst{\ref{schmidt}}, 
    Benoît Tabone\orcidlink{0000-0002-1103-3225}\inst{\ref{saclay}}, 
    Laurent Pueyo\orcidlink{0000-0003-3818-408X}\inst{\ref{stsci}}, 
    Marshall Perrin\orcidlink{0000-0002-3191-8151}\inst{\ref{stsci}}, 
    William O. Balmer\orcidlink{0000-0001-6396-8439}\inst{\ref{jhu},\ref{stsci}}, 
    Sarah Betti\orcidlink{0000-0002-8667-6428}\inst{\ref{stsci}}, 
    Christine H. Chen\orcidlink{0000-0002-8382-0447}\inst{\ref{stsci},\ref{jhu}}, 
    John H. Debes\orcidlink{0000-0002-1783-8817}\inst{\ref{stsci}}, 
    Julien H. Girard\orcidlink{0000-0001-8627-0404}\inst{\ref{stsci}}, 
    Kielan K. W. Hoch\orcidlink{0000-0002-9803-8255}\inst{\ref{stsci}}, 
    Jens Kammerer\orcidlink{0000-0003-2769-0438}\inst{\ref{eso}}, 
    Cicero Lu\orcidlink{0000-0001-9352-0248}\inst{\ref{gemini}}, 
    Isabel Rebollido\orcidlink{0000-0002-4388-6417}\inst{\ref{esa_madrid}}, 
    Emily Rickman\orcidlink{0000-0003-4203-9715}\inst{\ref{esa_baltimore}}, 
    Connor Robinson\orcidlink{0000-0003-1639-510X}\inst{\ref{Alfred}}, 
    Kadin Worthen\orcidlink{0000-0002-5885-5779}\inst{\ref{jhu}}, 
    Roeland P. van der Marel\orcidlink{0000-0001-7827-7825}\inst{\ref{stsci},\ref{jhu}}, 
    Nikole K. Lewis\orcidlink{0000-0002-8507-1304}\inst{\ref{cornell}}, 
    Sara Seager\orcidlink{0000-0002-6892-6948}\inst{\ref{mit_kavli}, \ref{mit_2}, \ref{mit_3}}, 
    Jeff A. Valenti\orcidlink{0000-0003-3305-6281}\inst{\ref{stsci}}, %
    Remi Soummer\orcidlink{0000-0003-2753-2819}\inst{\ref{stsci}}}
    
\institute{Department of Physics \& Astronomy, Johns Hopkins University, 3400 N. Charles Street, Baltimore, MD 21218, USA\label{jhu}
\and Space Telescope Science Institute, 3700 San Martin Drive, Baltimore, MD 21218, USA\label{stsci}
\and Department of Astronomy, Smith College, Northampton, MA, 01063, USA\label{smith}
\and Earth and Planets Laboratory, Carnegie Institution for Science, 5241 Broad Branch Road, NW, Washington, DC 20015, USA\label{carnegie}
\and Astrophysics \& Space Institute, Schmidt Sciences, New York, NY 10011, USA\label{schmidt}
\and Université Paris-Saclay, CNRS, Institut d’Astrophysique Spatiale, 91405 Orsay, France\label{saclay}
\and European Southern Observatory, Karl-Schwarzschild-Straße 2, 85748 Garching, Germany\label{eso}
\and Gemini Observatory/NSF NOIRLab, 670 N. A’ohoku Pl., Hilo, HI 96720, USA\label{gemini}
\and European Space Agency (ESA), European Space Astronomy Centre (ESAC), Camino Bajo del Castillo s/n, 28692 Villanueva de la Ca\~nada, Madrid, Spain\label{esa_madrid}
\and European Space Agency (ESA), ESA Office, Space Telescope Science Institute, 3700 San Martin Dr, Baltimore, MD 21218, USA\label{esa_baltimore}
\and Division of Physics and Astronomy, Alfred University, 1 Saxon Drive, Alfred, NY 14802, USA\label{Alfred}
\and Department of Astronomy and Carl Sagan Institute, Cornell University, 122 Sciences Drive, Ithaca, NY 14853, USA\label{cornell}
\and Department of Physics and Kavli Institute for Astrophysics and Space Research, Massachusetts Institute of Technology, Cambridge, MA 02139, USA\label{mit_kavli}
\and Department of Earth, Atmospheric and Planetary Sciences, Massachusetts Institute of Technology, Cambridge, MA 02139, USA\label{mit_2}
\and Department of Aeronautics and Astronautics, MIT, 77 Massachusetts Avenue, Cambridge, MA 02139, USA\label{mit_3}
\and NASA Goddard Space Flight Center, 8800 Greenbelt Road, Greenbelt, MD 20771, USA\label{goddard}
\and Sellers Exoplanets Environment Collaboration, 8800 Greenbelt Road, Greenbelt, MD 20771, USA\label{seec}
\and Astrophysics Division, Science Mission Directorate, NASA Headquarters, 300 E Street SW, Washington, DC 20546, USA\label{nasa_hq},
\and Association of Universities for Research in Astronomy, 1331 Pennsylvania Avenue NW Suite 1475, Washington, DC 20004, USA\label{aura}}

\date{}
    \abstract
   {Young accreting planetary-mass objects are thought to draw material from a circumplanetary disk composed of gas and dust. While the gas within the disk is expected to disperse within the first million years, strong accretion has nonetheless been detected in older systems, including the 30–45 Myr-old planetary-mass companion Delorme\,1\,AB\,b.}
   {We conducted spectroscopic observations with the James Webb Space Telescope's Mid-Infrared Instrument (\textit{JWST}/MIRI) to investigate the presence of circumplanetary material around this young accreting planet and to characterize the planet's atmospheric properties and composition.}
   {We performed forward modeling using atmospheric models to characterize the planet's atmosphere, combining our MIRI observations with archival ground-based near-infrared data.
   We used slab models to analyze the circumplanetary gas and investigated H$_2$ emission.}
   {We derived the atmospheric parameters of Delorme\,1\,AB\,b, finding an effective temperature of $T_{\mathrm{eff}} = 1725\pm134$\,K. 
   To achieve a satisfactory fit to the observed spectrum, 
   a secondary component is required, consistent with dust emission from a circumplanetary disk (CPD), 
   characterized by a blackbody temperature of $T_{\mathrm{bb}} = 295 \pm 27$\,K and an effective radius of $R_{\mathrm{bb}} = 18.8\pm2.7\,R_{\mathrm{Jup}}$. 
   Beyond $10\,\mu$m, the spectral energy distribution (SED) becomes dominated by this circumplanetary disk rather than the planet itself. 
   We detected strong emission from HCN and C$_2$H$_2$, along with tentative evidence of the isotopologue $^{13}$CCH$_2$,
   while no O-bearing species such as CO, CO$_2$, or H$_2$O are observed in the CPD spectrum.
   This suggests that the gas in the CPD has an elevated C/O. 
   We also identified spatially extended H$_2$ emission around the planet, tracing warm gas, with indications that it may be at a higher temperature than the non-extended component.}
   {The mid-infrared spectrum of the planetary-mass companion Delorme\,1\,AB\,b reveals the first detection of bright C-bearing species in a CPD together with an outflow traced by H$_2$ extended emission, which could be interpreted as a disk wind.
   The hot dust continuum emission suggests an inner cavity in the CPD.
   The presence of warm gas in the CPD provides constraints on the disk’s chemical composition and physical conditions, opening up new avenues for disk studies.
   The study of these long-lived ``Peter Pan'' disks will enhance our understanding of how accretion persists in evolved low-mass systems and shed light on their formation, longevity, and evolutionary pathways in planetary systems.}
\authorrunning{M. Mâlin et al.}
\titlerunning{Delorme 1\,AB\,b observed at mid-infrared wavelengths} 
\maketitle

\section{Introduction}
During the first million years of planetary formation, young planets are still forming and can be surrounded by gas and dust in the form of a circumplanetary disk (CPD), from which they accrete mass.
These disks play a critical role in regulating accretion and shaping the early evolution of the planet and the formation of potential moons.
A prime example is the planetary system PDS\,70, with the first directly imaged protoplanets embedded in their disks \citep{keppler_discovery_2018, mesa_vltsphere_2019}, which shows both spectroscopic signatures of accretion \citep{haffert_two_2019} and evidence of a surrounding CPD \citep{christiaens_evidence_2019, benisty_circumplanetary_2021}.
Similar characteristics have been observed in GQ\,Lup\,B, a 10--30M$_\mathrm{Jup}$ companion that shows compelling signs of active accretion \citep{neuhauser_evidence_2005, demars_emission_2023}, with recent mid-infrared (MIR) observations confirming the presence of circumstellar dust emission as well \citep{stolker_characterizing_2021, cugno_mid-infrared_2024}.
An increasing population of planetary-mass objects have shown evidence of circumplanetary materials through accretion markers or IR/sub-millimeter excess, for example, the protoplanet candidate AB Aur b \citep{currie_images_2022, bowler_h_2025, shibaike_predictions_2025}, wide companions such as Sr\,12\,c \citep{santamaria-miranda_accretion_2018, wu_alma_2022}, DH\,Tau\,b \citep{zhou_accretion_2014, lazzoni_search_2020}, GSC\,6214-210\,B \citep{van_holstein_survey_2021}
and YSES\,1\,b \citep{hoch_silicate_2025}, 
and free-floating objects such as OTS\,44 \citep{luhman_new_2008, joergens_ots_2013} and Cha\,1107-7626 \citep{luhman_discovery_2005, flagg_detection_2025}.
These planetary companions are located in young, forming systems less than 5 Myr old and are actively accreting material from their surrounding environments, except for YSES 1 b, which is estimated to be slightly older at 17 or 27 Myr \citep{zhang_eso_2024}.
Based on the observations cited above,
there appears to be a trend from younger objects, with a rising spectral energy distribution (SED) toward longer wavelengths with evidence of dust features, to more evolved systems exhibiting dust settling and flatter SEDs.
While some CPDs have been detected at sub-millimeter wavelengths with ALMA (e.g., PDS 70), others remain undetected in this regime despite exhibiting clear disk signatures — such as red near-infrared colors, mid-infrared excess, or spectroscopic evidence of ongoing accretion \citep{wu_alma_2020}. 
Similarly, accretion signatures including the near-infrared (NIR) hydrogen series have also been observed in young Class 0/I proto-brown dwarfs \citep{riaz_accretion_2021}, albeit with 3-4 dex higher estimated accretion rates ($\dot{M}_\mathrm{acc}\sim10^{-6}-10^{-8} M_{\odot}$/yr).
Substellar systems offer invaluable laboratories for probing the accretion physics and disk-planet interactions at play during the earliest stages of planetary evolution. 

Among these young and accreting planetary-mass objects, Delorme\,1\,AB\,b \citep[2MASS J01033563-5515561(AB)b,][]{delorme_direct-imaging_2013} stands out due to its relatively older age of 30--45 Myr as a member of the Tucana-Horologium association.
This companion has an estimated mass of $13 \pm 5$~M$_\mathrm{Jup}$ and orbits at a projected separation of 84~au (astronomical units) around a close binary system composed of two M\,5.5 type stars separated by 12~au and located at a distance of 47.2~pc (parsec) \citep{gaia_collaboration_gaia_2023}.
The SED of Delorme\,1\,AB\,b, from the UV to visible wavelengths, is consistent with a spectral type of $\mathrm{L}0\pm0.5$ at an effective temperature of $\sim$ 2000~K \citep{luhman_jwstnirspec_2023, janson_binaries_2017}.
The detection of strong absorption lines (e.g., Na{\sc i}, CrH, FeH, VO) indicates a low-gravity atmosphere typical of an object younger than 100\,Myrs \citep{eriksson_strong_2020}.
Its spectrum, obtained with the VLT/MUSE and the UVES spectrograph,
reveals numerous hydrogen-recombination lines, including H${\alpha}$ and H${\beta}$, as well as weak He{\sc i} emissions \citep{eriksson_strong_2020, ringqvist_resolved_2023}.
The NIR spectrum also reveals Pa$\beta$, Pa$\gamma$, and Br$\gamma$ emission lines \citep{betti_near-infrared_2022}.
These emission lines trace accretion from circumplanetary material. 
At the estimated age of the system, standard evolutionary models predict that the surrounding gas should have already dispersed. 
Typical disk lifetimes are known to be only a few million years \citep[e.g.][]{pecaut_star_2016}, so this unexpected accretion signature may indicate that the system is younger than previously thought.
However, recent discoveries of long-lived, actively accreting disks around very low-mass stars and brown dwarfs — referred to as Peter Pan disks — have challenged conventional assumptions about disk dispersal timescales.
Proposed explanations include giant planetesimal collisions \citep{flaherty_planet_2019}, star–disk dynamical interactions, and tidal disruptions by giant planets \citep{silverberg_peter_2020}, though it remains unclear why such scenarios would be more common for low-mass stars and brown dwarfs.
The nature and persistence of long-lived disks remain poorly understood, highlighting the need for further observational and theoretical investigations.

The luminosity of the accretion lines observed in Delorme\,1\,AB\,b implies a relatively high mass accretion rate, estimated between $0.2$ and $5 \times 10^{-8}$ M$_\mathrm{Jup}$/yr 
\citep[depending on the formalism used to scale the line luminosities with accretion;][]{betti_near-infrared_2022}. 
For comparison, typical mass accretion rates for planetary-mass objects are in the range $10^{-6}$–$10^{-8}$\,M$_\mathrm{Jup}$/yr \citep{zhou_accretion_2014}.
The inferred values for Delorme\,1\,AB\,b are more consistent with models of planetary shock accretion \citep{aoyama_theoretical_2018}, although magnetospheric accretion at the planet’s surface cannot be excluded \citep{bouvier_magnetospheric_2007, zhu_accreting_2015}.
The high accretion rate observed in Delorme\,1\,AB\,b strongly suggests the presence of a circumplanetary disk, whose reservoir of gas would be necessary to maintain such high accretion at this age.

The companion Delorme\,1\,AB\,b poses a significant challenge to current planet formation theories. 
Its estimated mass is too low to be readily explained by stellar gravitational instability, yet its wide orbital separation renders formation via core accretion highly unlikely \citep{delorme_direct-imaging_2013}.
Formation scenarios for circumbinary planets remain especially uncertain, and no current model can simultaneously account for its observed mass, separation, and ongoing accretion activity. 
Recent work suggests that these properties may be better reproduced by invoking outward migration or dynamical scattering influenced by the central binary \citep{teasdale_potential_2024}. 
This object represents a compelling case study for refining our understanding of the processes governing the formation and evolution of planetary-mass companions in complex stellar environments.\\

Mid-infrared observations are especially valuable for assessing disk properties.
\textit{Spitzer} spectroscopy has demonstrated the power of the MIR in characterizing T~Tauri protoplanetary disks, particularly their molecular emission \citep{lahuis_hot_2006, carr_organic_2008, pontoppidan_spitzer_2010, salyk_spitzer_2011}
as well as tracers of disk winds such as atomic [Ne\,\textsc{ii}] lines \citep{pascucci_detection_2007} and molecular H$_2$ lines \citep{nomura_molecular_2005}.
MIR spectroscopy with \textit{Spitzer} has also been used to study slightly older debris disks and their dust features \citep[e.g.][]{lu_trends_2022}.
Recently, the advent of MIR spectroscopy from 4.9 to 27.9\,$\mu$m with the \textit{JWST} MIRI medium-resolution spectrometer (MRS) has opened up a new window for studying circumstellar and circumplanetary environments at unprecedented sensitivity and spectral resolution (R$_\lambda$~$\sim$~3700). 
An increasing number of emission lines are subsequently appearing in disk spectra, in particular around very low-mass \citep[e.g.][]{arabhavi_minds2_2025} and T Tauri stars \citep[e.g.][]{banzatti_water_2025}.
These observations have revealed a rich chemistry in young disks, including evidence of carbon-rich environments around very low-mass stars (VLMS; 
\citealt{arabhavi_abundant_2024, arabhavi_minds_2025,
kanwar_minds_2024,
xie_water-rich_2023}) and brown dwarfs \citep[BDs;][]{perotti_minds_2025,
morales-calderon_minds_2025},
as well as the presence of numerous molecules and hydrocarbons in T Tauri disks \citep{perotti_water_2023, colmenares_jwstmiri_2024, 
grant_minds_2024, vlasblom_minds_2025, 
temmink_minds_2025,
arulanantham_jdisc_2025, ramirez-tannus_xue_2025}.
The growing number of observations across different objects is beginning to reveal emerging trends \citep{grant_transition_2025, arabhavi_minds2_2025, 
arulanantham_jdisc_2025, banzatti_water_2025, van_dishoeck_diverse_2023}.
The chemical composition of disks surrounding objects less massive than T Tauri and very low-mass stars remains unexplored, and MIRI/MRS offers the sensitivity needed to enable the first in-depth characterization of disks around planetary-mass companions.
For example, \citet{flagg_detection_2025} report evidence of hydrocarbon emission in one such system but also illustrate that with the MIRI low resolution spectrometer (LRS; R\,$\sim$\,100), it remains challenging to constrain the gas properties. 
This further highlights the added value of the higher spectral resolution provided by MIRI/MRS.

In addition to the molecular signatures, MIR emission lines such as H$_2$, [Ar\textsc{ii}], and [Ne\textsc{ii}] have been detected, tracing disk winds and outflows \citep[see, e.g.][]{espaillat_jwst_2023}.
By leveraging the integral field unit capabilities of MIRI, extended emission features can be spatially resolved, allowing for the detection of structures associated with disk winds and outflows, as highlighted in recent studies of young stellar objects \citep{delabrosse_jwst_2024, bajaj_jwst_2024, arulanantham_jwst_2024, schwarz_minds_2025}.
The objects studied in these recent works, with typical masses ranging from 0.1 to 2\,M$_{\odot}$ and luminosities between 0.2 and 2\,L$_{\odot}$, are all more massive and luminous 
than the 13 $M_{\rm{Jup}}$ (0.013\,M$_{\odot}$) planetary-mass companion presented in this paper.

Our study provides the first detailed MIR investigation of a circumplanetary disk, 
specifically that surrounding the companion Delorme\,1\,AB\,b.
In Sect.~\ref{sec:data_red}, we outline the observations and the data reduction procedures applied to obtain the spectra.
The primary goals of this study are twofold: to characterize the atmosphere of Delorme\,1\,AB\,b (Sect.~\ref{sec:atm_charact}) and characterize the CPD surrounding the companion (Sect.~\ref{sec:cpd}). 
We present the interpretation of our findings in Sect.~\ref{sec:discussion} and summarize our conclusions in Sect.~\ref{sec:conclusion}.

\section{Observations and data reduction}
\label{sec:obs_data_red}
\subsection{Program observations}
Observations of Delorme\,1\,AB were carried out with the \textit{JWST}/MIRI Medium Resolution Spectrometer \citep[MRS,][]{wells_mid-infrared_2015, argyriou_jwst_2023}
by the \textit{JWST} Telescope Scientist Team (JWST-TST)\footnote{\url{https://www.stsci.edu/~marel/jwsttelsciteam.html}}.
This collaboration uses Guaranteed Time Observations (GTO, PI: M. Mountain) across three science topics: Exoplanet and Debris Disk High-Contrast Imaging (lead: M. Perrin), Transiting Exoplanet Spectroscopy (lead: N. Lewis), and Local Group proper-motion Science (lead: R. van der Marel).
Previous studies from the TST High-contrast series includes \citet{rebollido_jwst-tst_2024, kammerer_jwst-tst_2024, ruffio_jwst-tst_2024, hoch_jwst-tst_2024, balmer_jwst-tst_2025}.
The observations studied in this work are part of the GTO Program 2778 (PI: M. Perrin).
It provides integral field spectroscopy covering wavelengths from 4.9 to 27.9 $\mu$m, divided into four co-axial channels.
A four point-source positive dither pattern was used to improve PSF spatial sampling \citep{wells_mid-infrared_2015}.
Due to its reduced sensitivity, channel 4 (from 17.9 to 27.9 $\mu$m) was not used at its full resolution in this analysis.
The exposure parameters were set to 100 groups for a single integration, and  
the total exposure time was 1110 seconds per MRS sub band. 
We use the FAST readout mode of the detector.
The observation strategy was designed according to recommendations from ERS\,1386 \citep{hinkley_jwst_2022}, similar to the approach used for the isolated planetary-mass companion VHS\,1256\,b \citep{miles_jwst_2023}, and did not include dedicated background exposures.

\subsection{Data reduction}
\label{sec:data_red}
The data were retrieved from the Mikulski Archive for Space Telescopes \citep[MAST,][]{marston_overview_2018} archive, already processed through the \texttt{Detector1Pipeline} to produce rate files.
The \texttt{Spec2Pipeline} was then applied with the default MRS steps, excluding the residual fringing correction \citep[as recommended for sources with dense molecular bands][]{gasman_jwst_2023}.
Spectral cubes were constructed using the \texttt{Spec3Pipeline} with the drizzle algorithm \citep{law_3d_2023} in \texttt{IFUALIGN} mode to minimize interpolation artifacts in the cube construction,
and we enabled the outlier rejection step.
The data were processed through all three stages of the \texttt{JWST} pipeline \citep{bushouse_jwst_2025} version 1.15.1, using the CRDS reference context 1293.

Background contribution must be removed, as it becomes a non-negligible source of brightness at wavelengths beyond 12 $\mu$m.
The background emission is estimated at each wavelength by measuring the histogram of the pixels within the field of view.
We fit the histogram with a Gaussian function to determine its peak,  corresponding to the background level.
This background value is then subtracted from each cube image to produce background-corrected data cubes, under the assumption that the background emission is uniform across the field of view.

\begin{figure*}
    \centering
    \includegraphics[width=1\linewidth]{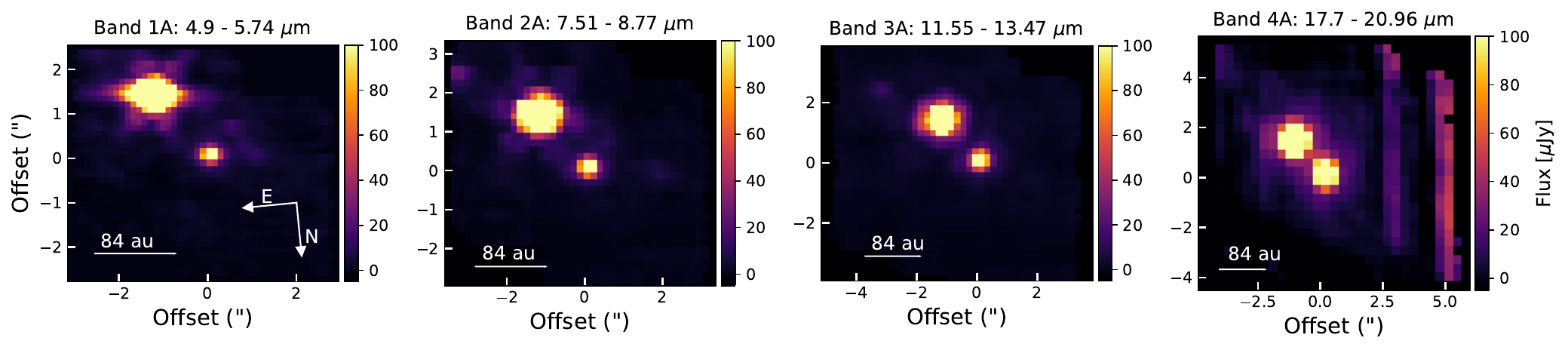}
    \caption{Median of the cubes for band A across all channels after removing the background contribution.
    The planet Delorme\,1\,AB\,b is in the middle of the field of view,
    while the point source in the upper corner represents the binary Delorme\,1\,AB.
    The vertical stripes observed in band 4A are caused by detector artifacts.}
    \label{fig:cubes}
\end{figure*}

Figure \ref{fig:cubes} display the final background-subtracted median cubes for band A.
The positions of the host binary star and its planetary-mass companion are determined by performing a 2D Gaussian fit on these median cubes for each band.
The flux of both objects is then extracted in an aperture of the radius of the full width at half maximum (FWHM) of the MIRI/MRS point spread function (PSF) for each wavelength \citep{argyriou_jwst_2023}.
We applied an aperture correction factor at each wavelength to account for the finite aperture \citep{law_james_2025}, and a $\sigma$-clipping with $\sigma = 10$ to account for any potential outliers.
The uncertainties provided by the \texttt{JWST} pipeline can be underestimated.
Therefore, we calculated the standard deviation of the flux in multiple apertures, positioned at the same separation between the host and the planet at different position angle spaced by 20 degrees.
This uncertainty term is added in quadrature to the one from the \texttt{JWST} pipeline.
This method is similar to the one used in \cite{kuhnle_water_2025}.
It increases the uncertainties at each wavelength by at least a factor 2, providing more realistic values.
Finally, we did not apply any flux scaling between MRS spectral bands, as no significant offsets are present in this high signal-to-noise ratio ($S/N$) spectrum.

\subsection{Qualitative description of the spectrum}
\begin{figure*}
    \centering
    \includegraphics[width=1\linewidth]{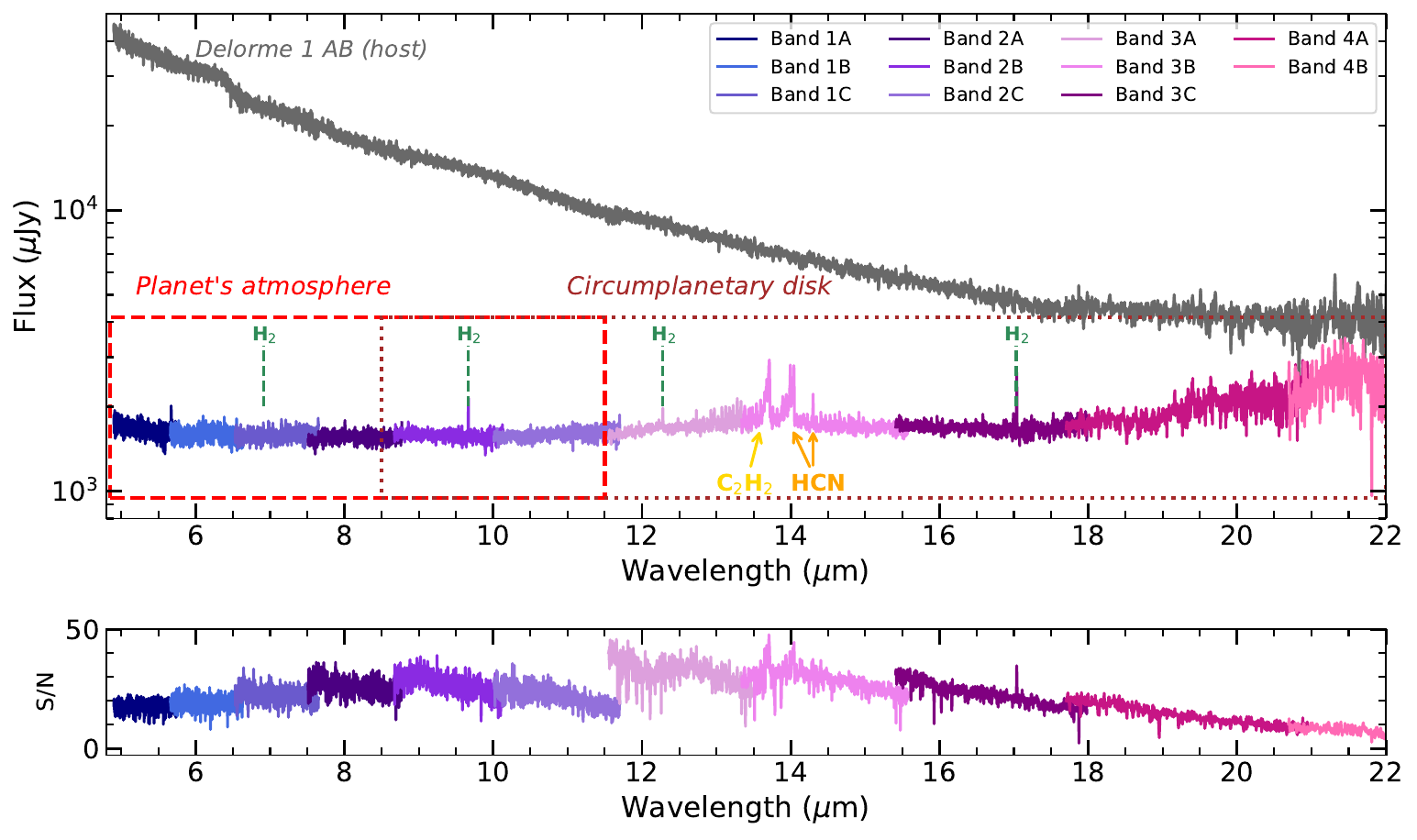}
    \caption{Top: Extracted spectra of the binary star (shown in gray, with the highest flux) and its companion Delorme\,1\,AB\,b (color coded from blue to purple, the lowest flux) across the MIRI/MRS bands. 
    The dashed region outlined in red indicates the wavelength range where the planet's atmosphere is detected (see Fig. \ref{fig:ccf_atm}), while the brown, dotted region marks the wavelengths dominated by the circumplanetary disk.
    These regions are shown for reference purposes and should not be considered as an actual boundary between the planet's atmospheric spectrum and the CPD.
    Molecular gases are shown for indicative purposes (see figure \ref{fig:slab_models} for detailed characterization).
    Bottom: $S/N$ as a function of wavelength for the planetary-mass companion.}
    \label{fig:spectra}
\end{figure*}

The extracted spectra are presented in Fig.~\ref{fig:spectra}, with flux displayed on a logarithmic scale as a function of wavelength. 
While a monotonic decrease in flux with increasing wavelength would be expected from the planet alone, approaching the Rayleigh-Jeans tail of the planet’s SED, the observed spectrum appears relatively flat, indicating a significant IR excess.
At longer wavelengths, the companion becomes nearly as bright as the binary host, with a flux ratio of only a factor of about three, using the median flux in band 3C.
This implies a very low contrast between the two objects at MIR wavelengths.
The spectral features observed in the first channel align well with expectations for a planetary atmosphere (see Sect. \ref{sec:atm_charact}).
The IR excess that emerges in channel 2 is consistent with emission from a CPD.
This CPD exhibits prominent molecular features, including the characteristic $Q-$branches of C$_2$H$_2$ and HCN near 14\,$\mu$m, both clearly identifiable (see Sect. \ref{sec:cpd_molec}). 
Numerous H$_2$ emission lines are also clearly visible (see Sect. \ref{subsec: molecular hydrogen}).
No silicate dust features are detected around 10\,$\mu$m, where such signatures would typically be expected. 
This absence of a prominent silicate feature suggests that the dust grains have grown to sizes greater than 5\,$\mu$m and settled \citep[cf.][]{kesslersilacci_c2d_2006, przygodda_evidence_2003,
tabone_rich_2023, arabhavi_abundant_2024}.
The dust is only detected through the disk’s continuum emission.
The channel 4 (17.7–27.9\,$\mu$m) spectra is not used in this analysis, as the $S/N$ per wavelength becomes low. Additionally, the PSFs of both objects begin to overlap and residual background contamination may be present.
Moreover, these wavelengths are dominated by the CPD component, which does not exhibit any of the features targeted for further investigation in this study.

\section{Atmospheric characterization of the planet}
\label{sec:atm_charact}

To characterize the planet's atmosphere, we fit grids of atmospheric models in Sect.~\ref{subsec:atm_modeling} and similarly for the host binary star in Sect.~\ref{subsec:binary_stars}. 
We perform a cross correlation analysis in Sect.~\ref{subsec:cc_analysis}, and assess the overall atmospheric properties in Sect.~\ref{subsec:planet_atm_param}.

\subsection{Atmospheric modeling}
\label{subsec:atm_modeling}
Our analysis includes the NIR spectrum from VLT/MUSE \citep{eriksson_strong_2020},
the TripleSpec spectrum in bands J, H and K \citep{betti_near-infrared_2022} and the NIR VLT/NACO photometry \citep{delorme_direct-imaging_2013}.
We use only the observations with the highest $S/N$ from the TripleSpec observation nights, and we rescale these to each of the VLT/NACO photometry points in J, H and K bands to account for variations in flux calibration across instruments.

We use \texttt{species} \citep{stolker_species_2023} to fit the observed spectrum with atmospheric models of pre-computed grids of models.
Among the atmospheric grids available for this temperature regime and at medium spectral resolution, we choose to run the atmospheric fit with 
\texttt{Exo-REM} \citep{charnay_self-consistent_2018, blain_1d_2021},
\texttt{ATMO} \citep[grid from][]{petrus_x-shyne_2023}, 
\texttt{Sonora diamondback} \citep{morley_sonora_2024}, and we compared with the commonly used \texttt{BT-Settl} \citep{allard_atmospheres_2012}.
These grids are all suited to characterize the atmosphere of young giant directly imaged planets.
In particular, we expect that cloudy models will provide a better fit for the atmosphere of an L-dwarf spectrum such as Delorme\,1\,AB\,b studied here.
The \texttt{BT-Settl} models are widely used for late M, L, and T dwarfs. 
It includes microphysical processes and settling to account for clouds.
The \texttt{Exo-REM} models focus on L and T dwarfs and include a simple microphysics cloud model and disequilibrium chemistry, which becomes important for late L and early T dwarfs as cooler temperatures begin to alter timescales of vertical mixing and chemical reactions in their atmospheres.
The \texttt{ATMO} models correspond to fully cloud-free atmospheres, but the lack of clouds is compensated invoking diabatic processes that change the temperature structure \citep{tremblin_fingering_2015, tremblin_thermo-compositional_2019}.
This grid assumes a modification of the temperature gradient with an effective adiabatic index $\gamma$.
Among the \texttt{Sonora} models, we choose the \texttt{Diamondback} which correspond to the cloudy warmer (L and early T type) brown dwarfs and planets. This grid is the first cloudy \texttt{Sonora} grid, which parametrized clouds with the f$_{\rm sed}$ parameters, as previously used by \citet{ackerman_precipitating_2001}. This takes into account sedimentation in the atmosphere by assuming a balance between turbulent diffusion and sedimentation in horizontally uniform cloud decks.
The parameters and bounds of these grids are indicated in Appendix in Tab. \ref{tab:param_models_atm}.

We focus exclusively on the spectra from channels 1 to 3, as channel 4 suffers from reduced sensitivity.
The spectrum in channel 4 (> 18\,$\mu$m) appears to be dominated by the CPD features, so it does not further constrain the planetary atmospheric models.
To optimize computational efficiency, we constrain the temperature range between 1500\,K and 2000\,K, and we limit the radius between 0.5 and 10\,R$_\mathrm{Jup}$.
Using these bounds, we perform the fit using the \texttt{multinest} algorithm, employing 1000 live points.

The extracted MIRI/MRS spectrum reveals a clear IR excess confirming the presence of the circumplanetary disk. 
To account for this excess, we incorporate a blackbody component into the fit, treating its radius and temperature as free parameters.
The effective temperature of the CPD is bounded between 10\,K and 2000\,K (the upper bound is chosen to be cooler than the maximum temperature of the atmospheric models), while the CPD radius is constrained between 2 and 1000\,$R_\mathrm{Jup}$.
This approach enables to characterize the circumplanetary disk.
However, this presents a simplistic view of the CPD emission, as it assumes a uniform temperature throughout the CPD and does not account for a radial temperature profile.
In reality, a temperature gradient is expected, with higher temperatures closer to the planet (See Sect. \ref{sec:cpd}), since the CPD is primarily heated by passive irradiation from the central object.
\begin{figure*}[ht]
    \centering
    \includegraphics[width=1\linewidth]{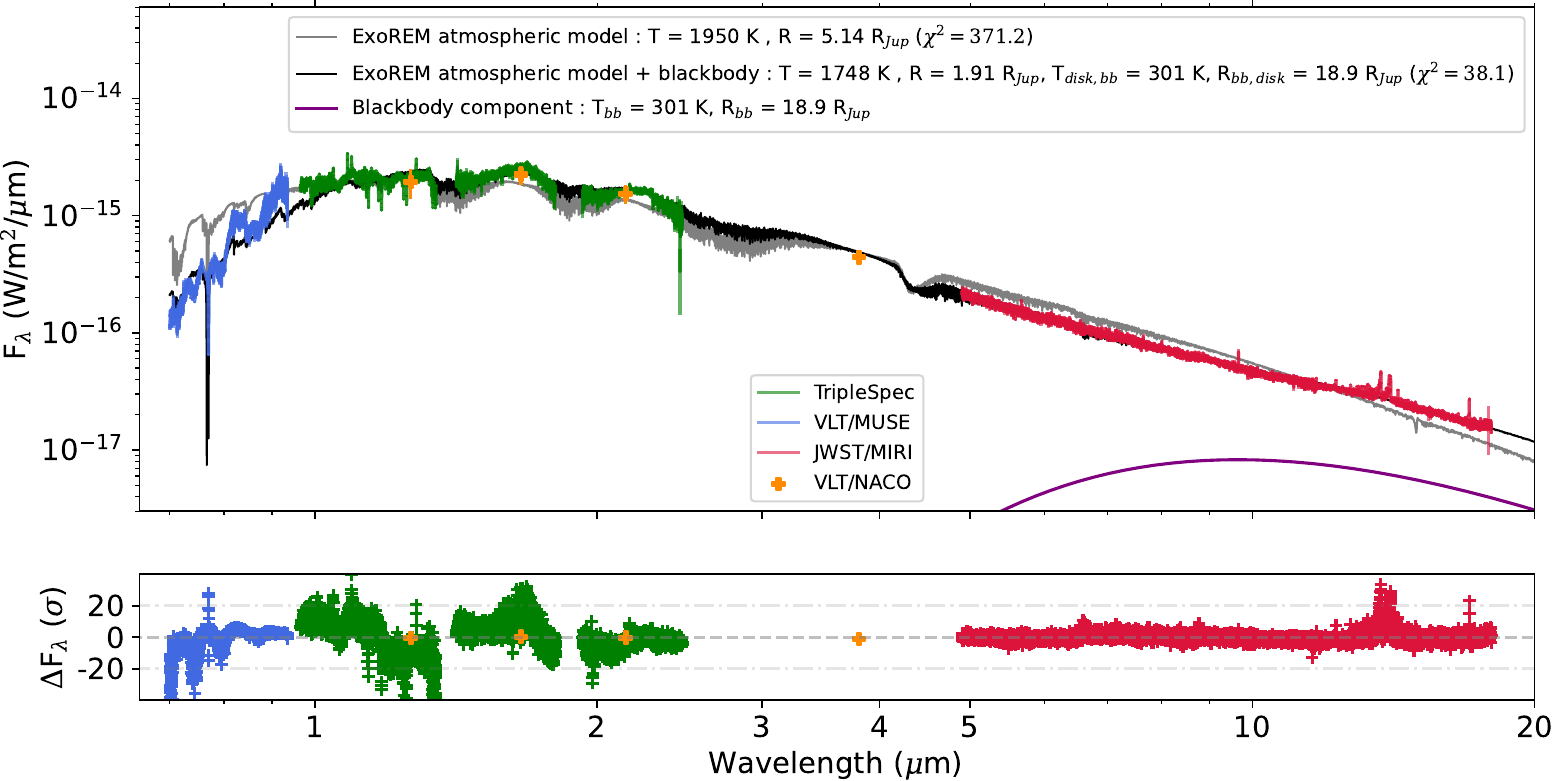}
    \caption{Delorme\,1\,AB\,b spectrum and photometric data from the visible to MIR along with the best-fit \texttt{Exo-REM} atmospheric models. The black curve includes an additional blackbody component to account for the observed MIR excess. More atmospheric model fits are available in Appendix~\ref{sec:appendix_atmos}.}
    \label{fig:best_fits_exorem}
\end{figure*}
%
The blackbody model with parameters that best represent the data has a temperature of 295$\pm$27\,K, 
with an effective area of 1111$\pm$318 R$_{\rm Jup}^2$,
if one assumes that the intensity is homogeneously distributed in the form of a disk. 
This is equivalent to a radius of 18.8$\pm$2.7\,R$_\mathrm{Jup}$.
Uncertainties are based on the weighted mean and standard deviation across the four atmospheric models: \texttt{Exo-REM}, \texttt{ATMO}, \texttt{Sonora}, \texttt{BT-Settl}.
Without this blackbody component, we note that the atmospheric models converge toward inconsistently low temperature or very large radii (e.g. 1500\,K and 2.9\,R$_{\rm Jup}$ with \texttt{ATMO} models, see Tab.\ref{tab:best_fits_atm_CPD}); thus, including the blackbody improves the fit to the data (from $\chi^2$ values, in Tab.\ref{tab:best_fits_atm_CPD}).
The lower limit for the planet's atmosphere at 1500\,K improves computational efficiency.
Without this constraint, 
the solution tends to converge to even more unphysically low temperatures.
In such cases, the model attempts to account for the MIR excess using only the atmospheric component, leading to a poor fit and unrealistically low planetary temperatures.
The atmospheric parameters derived from each model, with and without the inclusion of the blackbody component, are summarized in Table~\ref{tab:best_fits_atm_CPD}. Figure~\ref{fig:best_fits_exorem} presents an example in which we use \texttt{Exo-REM} models, providing the best-fit.

\begin{table*}[h!]
    \caption{Summary of the best-fit parameter measured with different atmospheric grids.}
    \centering
    \begin{tabular}{c|ccccccc:cc:c}
        \hline
        \hline
         Model & T$_{eff}$ & R (R$_\mathrm{Jup}$) & log(g) & [Fe/H] & C/O & f$_{sed}$ & $\gamma$ &
         T$_{bb}$ & R$_{bb}$ (R$_\mathrm{Jup}$) & $\chi^2$\\ 
         \hline
         \hline
         
         \texttt{Exo-REM} 
         & 1950$^{+1}_{-1}$ & 
         5.14$^{+0.2}_{-0.2}$ & 
         3.00$^{+0.01}_{-0.01}$ &
         2.00$^{+0}_{-0}$ &
         0.25$^{+0.01}_{-0.01}$ &
         -- &
         -- &
         -- & 
         -- &
         371.2\\

        \texttt{Exo-REM + BB} 
         & 1748$^{+2}_{-3}$ & 
         1.9$^{+0.1}_{-0.1}$ & 
         3.42$^{+0.03}_{-0.03}$ &
         1.49$^{+0.01}_{-0.01}$ &
         0.78$^{+0.01}_{-0.01}$ &
         -- &
         -- &
         301$^{+3}_{-5}$ & 
         18.9$^{+1}_{-1}$ &
         38.1\\

        \hline
         \texttt{ATMO} 
         & 1500$^{+0.1}_{-0.6}$ & 
         2.90$^{+0.12}_{-0.11}$ & 
         3.59$^{+0.05}_{-0.05}$ &
         -0.60$^{+0.01}_{-0.01}$ &
         0.70$^{+0.01}_{-0.01}$ &
         -- &
         1.04$^{+0.01}_{-0.01}$ &
         -- & 
         -- &
         288.7\\

         \texttt{ATMO + BB} 
         & 1860$^{+11}_{-11}$ & 
         1.66$^{+0.07}_{-0.07}$ & 
         2.61$^{+0.16}_{-0.08}$ &
         -0.32$^{+0.04}_{-0.07}$ &
         0.64$^{+0.04}_{-0.06}$ &
         -- &
         1.01$^{+0.01}_{-0.01}$ &
         311$^{+3}_{-3}$ & 
         17.79$^{+0.8}_{-0.7}$ &
         39.3\\

         \hline
         \texttt{Sonora} 
         & 1500$^{+0.1}_{-0.1}$ & 
         2.7$^{+0.1}_{-0.1}$ & 
         5.5$^{+0.01}_{-0.01}$ &
         -0.2$^{+0.02}_{-0.02}$ &
         -- &
         1.00$^{+0.01}_{-0.01}$ &
         -- &
         -- & 
         -- & 
         269.8\\

        \texttt{Sonora + BB} 
         & 1509$^{+8}_{-6}$ & 
         2.4$^{+0.1}_{-0.1}$ & 
         5.0$^{+0.03}_{-0.04}$ &
         0.43$^{+0.06}_{-0.03}$ &
         -- &
         1.28$^{+0.29}_{-0.24}$ &
         -- &
         259$^{+3}_{-3}$ & 
         23.3$^{+1}_{-1}$ & 
         59.1 \\

        \hline
        \texttt{BT-Settl} 
         & 1607$^{+1}_{1}$ & 
         2.87$^{+0.13}_{-0.12}$ & 
         3.00$^{+0.01}_{-0.01}$ &
         -- &
         -- &
         -- &
         -- &
         -- & 
         -- &
         466.6\\

        \texttt{BT-Settl + BB} 
         & 1689$^{+12}_{12}$ & 
         1.98$^{+0.09}_{-0.09}$ & 
         5.42$^{+0.03}_{-0.04}$ &
         -- &
         -- &
         -- &
         -- &
         321$^{+4}_{-4}$ & 
         17.0$^{+0.8}_{-0.8}$ &
         58.4\\
        \hline
    \end{tabular}
    \tablefoot{The second row for each model includes a blackbody component.}
    \label{tab:best_fits_atm_CPD}
\end{table*}

\subsection{Host stars}
\label{subsec:binary_stars}

The extracted spectrum of the binary host star is shown in gray in Fig.\,\ref{fig:spectra}. 
We find no evidence of accretion signatures or circumstellar material, consistent with previous findings from the NIR spectrum \citep{betti_near-infrared_2022}. 
The apparent MIR excess beyond 18\,$\mu$m is more likely due to background residuals and PSF overlap with the companion at longer wavelengths (see Fig.\,\ref{fig:cubes}).
We performed a simple forward modeling analysis of the binary host star using \texttt{species}, following the same approach as for the companion (see Sect.\ref{sec:atm_charact}).
The binary is unresolved with MIIR/MRS, and therefore we fit the spectrum considering both stars as a single component, assuming they have the same stellar properties. 
This provides an effective radius and mass values, accounting for the binary M dwarf - resulting in a higher $\log g$ than the true value.
Based on the \texttt{BT-Settl} model library, the best-fit parameters are a temperature of $T = 3524\,\mathrm{K}$, an effective radius of $R = 0.49\,R_\mathrm{\odot}$, and a surface gravity of $\log g = 5.5$.
Supporting figures are provided in the Appendix \ref{sec:appendix_atmos}, although we note that a detailed characterization of the binary host star is beyond the scope of this paper.

\subsection{Cross correlation analysis}
\label{subsec:cc_analysis}
\begin{figure}[ht]
    \centering
    \includegraphics[width=0.9\linewidth]{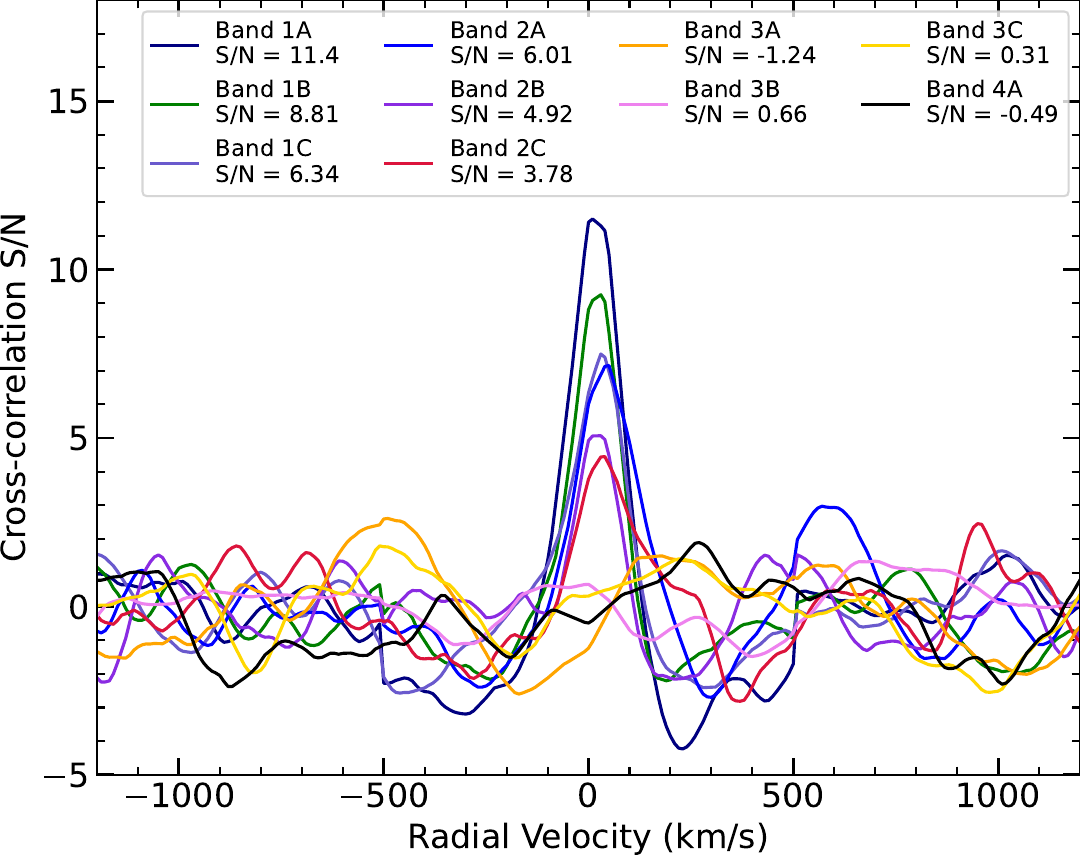}
    \caption{Cross correlation with \texttt{Exo-REM} model in all bands individually.
    The planetary atmosphere is detected until band 2C (wavelengths $\sim$ 10 $\mu$m).}
    \label{fig:ccf_atm}
\end{figure}

\begin{figure}[ht]
    \centering
    \includegraphics[width=0.9\linewidth]{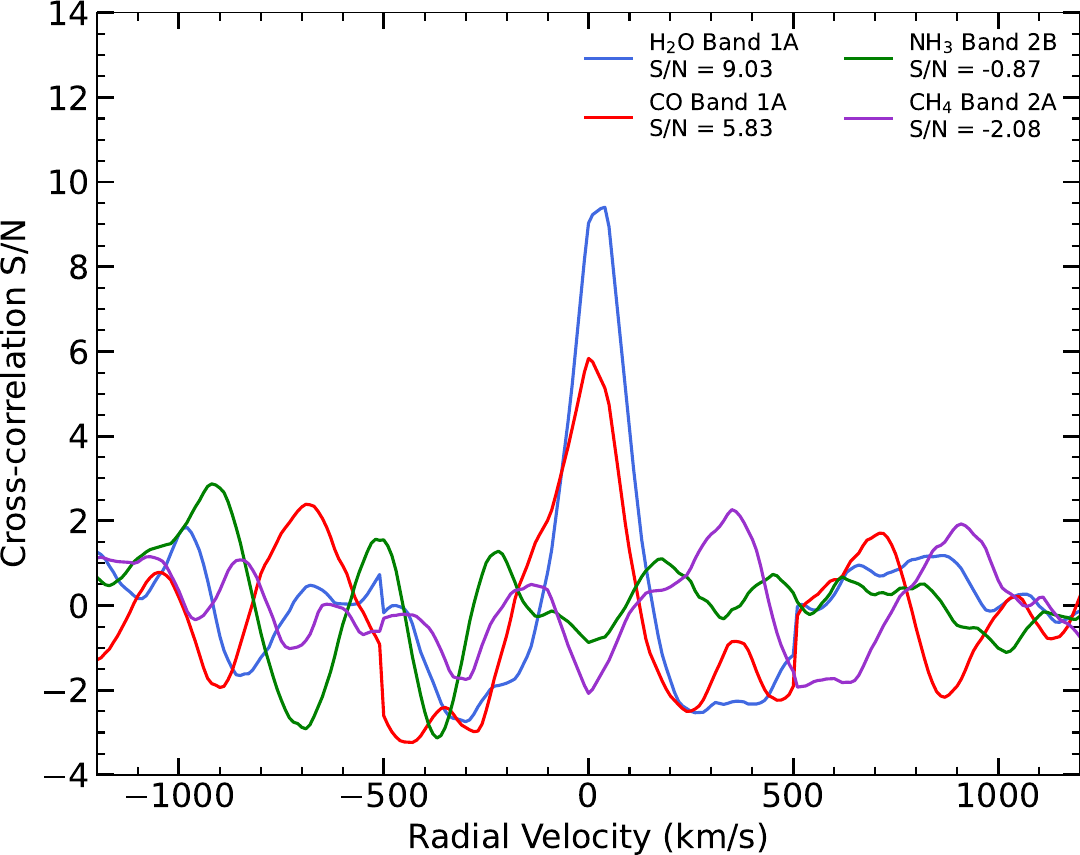}
    \caption{Cross correlation with \texttt{Exo-REM} molecular absorption templates for the planetary atmosphere.
    Strong detections of H$_2$O and CO are observed, while no significant signal is found for CH$_4$ or NH$_3$.}
    \label{fig:ccf_molec}
\end{figure}

Cross correlation is a powerful technique for detecting molecules in planetary atmospheres \citep[e.g.][]{konopacky_detection_2013, gandhi_jwst_2023}.
We applied it both to confirm molecular presence and to determine the wavelength range over which the planet's atmosphere is detected.
The observed MIRI spectrum was cross-correlated with the best-fit \texttt{Exo-REM} atmospheric model, revealing the planetary signal up to band 2C, approximately 10 $\mu$m.
The $S/N$ of the cross correlation function (CCF) peak ranged from $S/N$ = 11.4 in band 1A to $S/N$ = 3.78 in band 2C (Fig.~\ref{fig:ccf_atm}). 
Cross correlation with individual molecular absorption templates further confirmed the presence of CO ($S/N$ = 5.83) and H$_2$O ($S/N$ = 9) within band 1A, as depicted in Fig.~\ref{fig:ccf_molec}. 
The $S/N$ values were determined by dividing the CCF peak by the standard deviation of the CCF wings, measured after subtracting the autocorrelation signal over the velocity range of $\pm$500 to $\pm$3000 km/s. 
This analysis highlights the robust detection of key molecular species in absorption in the planet's atmosphere, and no other molecules from the planet's atmosphere are detected in the MIRI data.
This is consistent to what is expected in this temperature regime at mid-infrared wavelengths.
Figure \ref{fig:Spectrum_H2O_CO} in the Appendix shows that spectral features from these molecules can be directly seen in the companion's spectrum. Cross correlation enables to confirm their detection (mostly for CO that it more difficult to identify directly).


The planetary atmosphere dominates the spectrum in channel 1, while the spectral features weaken in channel 2 (7.51–11.7\,$\mu$m), although they remain detectable through cross correlation. 
The planet is not detected with a planet atmospheric model in channel 3.
This confirms that at wavelengths longer than those of channel 2, the spectrum is dominated by spectral features that differ entirely from an atmospheric model.

For comparison, we performed the same analysis on the host star's spectrum. We cross-correlated its spectrum with the best-fit model (see Sect. \ref{subsec:binary_stars}).
The cross correlation functions between the MRS spectrum of the host star and this model show a peak in all bands, with a $S/N$ ranging from 5 to 12, depending on the bands and their spectral features, which cause variations in the $S/N$.
We observe an offset in the radial velocity of the host binary, ranging from +70 to +100 km/s depending on the bands. 
The uncertainty in these values likely arises from line undersampling or wavelength calibration.

Although the contrast between the companion and the host is not particularly challenging and does not hinder direct spectral extraction, we also present an alternative approach for molecular detection, known as the ``molecular mapping'' method, as previously applied to MIRI/MRS simulated data \citep{patapis_direct_2022, malin_simulated_2023}. 
This complementary analysis is detailed in Appendix~\ref{sec:molecular_mapping}.
This method confirms the presence of H$_2$O and CO in both the planetary-mass companion and the host stars.
The correlation with hydrocarbon models also reveals the companion, demonstrating that the ``molecular mapping'' approach is effective at detecting circumplanetary disks.

\subsection{Planet's atmospheric parameters}
\label{subsec:planet_atm_param}
Overall, all atmospheric models provide generally consistent physical parameters. However, differences in cloud treatment across models account for variations in the measured parameters.
The atmospheric models \texttt{BT-Settl} and \texttt{Sonora diamondback} suggest lower temperatures and higher surface gravities, corresponding to unrealistically high mass values for the given radius
(with the mass measured from the atmospheric best-fits as : $M = gR^2/G$, g is the surface gravity, G the gravitational constant, R the radius of the planet).
From the $\chi^2$ values, \texttt{Exo-REM} and \texttt{ATMO} yield the best results ($\chi^2_{\rm red}$ = 38.1 and 39.3, respectively), which may reflect differences in their cloud parameterizations that influence how well they reproduce the emergent spectra.
We measure a super-solar C/O ratio, and metallicity values that depend on the atmospheric model considered. 
The model \texttt{Exo-REM} including the blackbody provides C/O = 0.78 and [Fe/H] = 1.49; while
\texttt{ATMO} suggests subsolar [Fe/H] = -0.32, and supersolar C/O = 0.64.
For the age and luminosity measured from the data of this planet, we expect a radius from 1.36  -- 1.49 R$_\mathrm{Jup}$ (older age assumption of 40\,Myrs) to 1.50 -- 1.55 R$_\mathrm{Jup}$ (younger age assumption of 30\,Myrs) based upon the \texttt{ATMO} evolutionary models and \texttt{Sonora} isochrones (Fig.~\ref{fig:evolution}). 
The radius is expected to decrease with age as the planet contracts, resulting in slightly larger radii at younger ages; however, these values remain closely aligned.
We also note that the ongoing accretion can slow down the contraction of the planet by bringing entropy \citep[e.g.][]{berardo_evolution_2017}.
%
Compared to our results, the \texttt{ATMO} atmospheric model provides a more consistent radius, while other models predict larger radii than expected from evolution models.
The object's mass cannot be determined from its orbit due to its long orbital period of approximately 1280 years \citep{delorme_direct-imaging_2013}.
However, based on evolutionary models that account for the system's age and measured luminosity, the estimated mass is 13.6–14 M$_\mathrm{Jup}$ if the system is 30\,Myrs. 
Isochrones for older ages suggest minimum masses of 13.8 M$_\mathrm{Jup}$ at 35 Myr and 14.5 M$_\mathrm{Jup}$ at 45 Myr, but they also allow for higher-mass solutions, reaching up to approximately 26 M$_\mathrm{Jup}$ at 45 Myr.
Furthermore, we measure nonphysical $\mathrm{log}$ g values — as low as 2.6\,dex with \texttt{ATMO}, even when the blackbody component is included.
Typically, $\mathrm{log}$ g is better constrained in the H-band due to the effects of collisionally induced absorption (CIA) from atmospheric H$_2$, particularly when combined with pressure effects on alkali line strengths \citep{kirkpatrick_discovery_2006}.
However, upon closer examination of the H-band, none of the models provide an optimal fit to the data, with \texttt{Sonora} performing slightly better due to its higher $\mathrm{log}$ g measurement.
When using only NIR data, the fit does not improve significantly, as other unphysical values emerge (lower temperature), though the H-band is somewhat better reproduced.

As shown in \cite{petrus_jwst_2024}, model fits over different wavelength ranges can yield varying results. 
Given the broad spectral coverage, it is not possible to robustly constrain all parameters simultaneously. This highlights the limitations of current models, which are not yet capable of accurately fitting data over such a wide wavelength range.

Moreover, we perform the fit while incorporating priors on radius and mass, as predicted by evolutionary models, i.e. $14\pm0.5 M_\mathrm{Jup}$ for the mass and $1.5\pm0.1 R_\mathrm{Jup}$ for the radius.
This approach yields the values presented in Table~\ref{tab:best_fits_atm_CPD_priors}.
The radius and mass values obtained from the atmospheric fit are indeed more consistent, but we note that these results are strongly influenced by these priors. 

As another test, we perform the atmospheric fitting using only the H-band.
The models \texttt{Exo-REM} and \texttt{ATMO}  provide values consistent with evolution models, but with higher temperatures. 
(\texttt{ATMO}: $R = 1.48\pm0.1 R_\mathrm{Jup}$, $\mathrm{log}g = 4.2\pm0.4$, $M =14^{+17}_{-9} M_\mathrm{Jup}$ and $T = 1960\pm50\,K$;
and \texttt{Exo-REM} $R = 1.53\pm0.1 R_\mathrm{Jup}$, $\mathrm{log}g = 4.77\pm0.3$, $M =56\pm26 M_\mathrm{Jup}$; and $T = 1970\pm40\,K$).
However, the mass values remain highly unconstrained by the atmospheric analysis.
These results emphasize the need to improve the spectral fitting of exoplanets using models that incorporate both atmospheric and evolutionary processes \citep[e.g.][]{wilkinson_breaking_2024}.

\section{Characterization of the circumplanetary disk}
\label{sec:cpd}
\subsection{Molecular gas emission}
\label{sec:cpd_molec}

Subtracting the circumplanetary disk dust continuum from the 1-D spectrum is crucial for accurately analyzing the molecular gas emission.
Continuum-subtraction is done using the method of \cite{temmink_minds_2024}, where the continuum is determined via an iterative fitting using a Savitzky-Golay filter with a third-order polynomial. 
Emission lines exceeding 2$\sigma$ above the continuum are masked to not skew the continuum estimation. 
The continuum is then subtracted, and all downward spikes more than 3$\sigma$ below the continuum are masked.
Finally, the baseline is determined using PyBaselines \citep{erb_pybaselines_2022} with the algorithm irsqr (Iterative Reweighted Spline Quantile Regression). There is a pseudo-continuum from the optically thick C$_2$H$_2$ component, so we exclude 6.8 to 8.25 $\mu$m and 10.5 to 17 $\mu$m from the continuum fit to accurately determine the gas properties. 
Details of the continuum estimation in the presence of this pseudo-continuum are provided in the Appendix \ref{fig:Delorme1ABb_CPD_C2H2_continuum}.

Molecular emission from C$_2$H$_2$, HCN, and $^{13}$CCH$_2$ are detected in the MIR spectrum of the CPD surrounding the young planet Delorme\,1\,AB\,b. We use local thermodynamic equilibrium (LTE) slab models, generated using molecular line data, to reproduce molecular emission visible in the spectrum (see \citealt{tabone_rich_2023} for further details).
This method has been successfully used to reproduce the emission from disks observed with JWST MIRI/MRS, including in systems with low-mass hosts \citep[e.g.][]{tabone_rich_2023, grant_minds_2023}.
The model includes three free parameters: the line-of-sight column density $N$, the gas temperature $T$, and the effective emitting area given by $\pi$$R$$^{2}$ for a disk of emission with radius $R$. 
We note that the parameter $R$ does not necessarily correspond to the radius of a disk, but may instead represent a ring with the same effective emitting area.
The model spectrum is convolved to a resolving power that matches the observations and resampled to have the same wavelength sampling as the observed spectrum.
We fit the spectral region from 13 to 15 $\mu$m which exhibits the molecular lines from the CPD.
The molecular fitting is done using an iterative approach where one molecular species is fit, the best-fit model is subtracted off, and then another molecular species is fit. This method has been found to produce results that are consistent with methods which fit all species simultaneously (e.g., \citealt{grant_minds_2023} and \citealt{kaeufer_bayesian_2024}). 
First, we fit the main C$_2$H$_2$ $Q-$branch from 13.6 to 13.72 $\mu$m. The best-fit is then subtracted to remove individual $P-$ and $R-$branch lines, and then the molecular pseudo-continuum is fit in the wings of the emission. This optically thick component is then subtracted before re-fitting the main $Q-$branch, which is borderline optically thin. The two C$_2$H$_2$ components are then removed and HCN and $^{13}$CCH$_2$ are fit. After inspecting the residuals after subtracting the total model, no additional molecular species are identified, except for H$_2$, which is analyzed in Section~\ref{subsec: molecular hydrogen}.
The best-fit model parameters are provided in Table \ref{tab:slab_spectrum} and these model spectra are compared to the observations in Figure \ref{fig:slab_models}  (see Fig\,\ref{fig:Delorme1ABb_CPD_C2H2_continuum} in Appendix for a broader wavelength coverage).  
The corresponding $\chi^{2}$ maps are provided in the Appendix in Sect. \ref{sec:chi2maps}.
\begin{figure*}
    \centering
    \includegraphics[width=18cm]{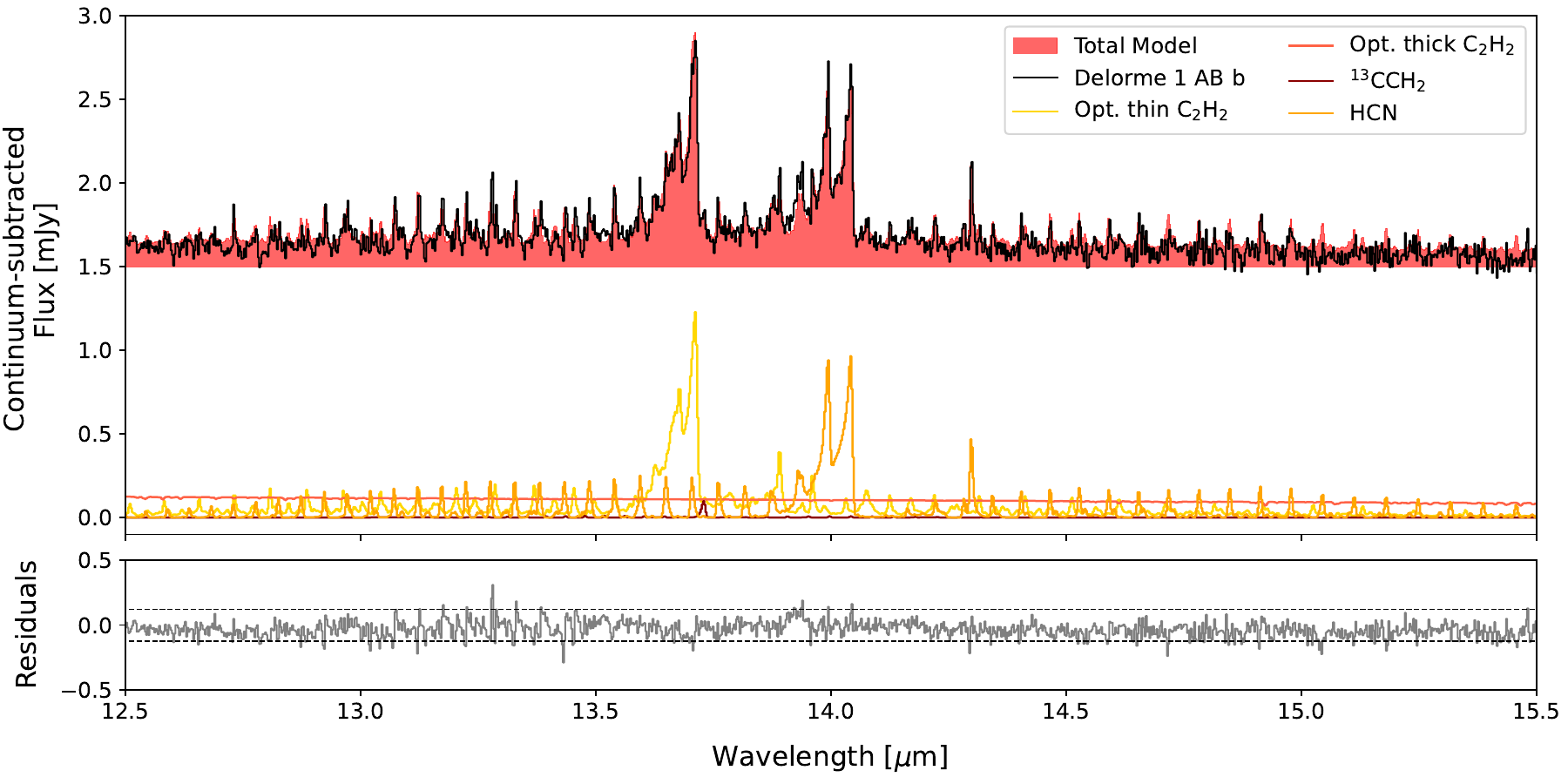}
    \caption{Modeling of the CPD gas.
    Top: Continuum-subtracted MIRI spectrum (black) compared to the best-fit total model (red).
    Emission from C$_2$H$_2$ is shown in yellow for the thin component and dark orange for the thick component, 
    HCN is in orange, and $^{13}$CCH$_2$ in dark red.
    Bottom: Residuals. Black dashed lines shows $\pm$ 3$\sigma$, for reference.}
    \label{fig:slab_models}
\end{figure*}

\begin{table}[ht]
    \centering
     \caption{Best-fit slab model results for molecules present in Delorme\,1\,AB\,b CPD.} 
    \begin{tabular}{ccccc}
    \hline
    \hline
        Species & T & N & R & $\mathcal{N}_{tot}$ \\
                &  [K] & [cm$^{-2}$] & [au] & [molecules]\\
        \\
        \hline
        C$_2$H$_2$ (thin)& 700 & 1.0$\times$10$^{17}$ & 0.0053 & 1.96$\times$10$^{39}$\\
        $^{13}$CCH$_2$ & 100 & 1.0$\times$10$^{14}$ & 0.788 & 4.36$\times$10$^{40}$ \\
        HCN & 450 & 2.15$\times$10$^{17}$ & 0.0094 & 1.34$\times$10$^{40}$\\
        \hline
    \end{tabular}
    \tablefoot{While we provide the best-fit values for $^{13}$CCH$_2$, we note that the molecular line list for this isotopologue may be incomplete and therefore the values should be taken with caution.  }
   \label{tab:slab_spectrum}
\end{table}

An optically thick C$_2$H$_2$ component is required to fit the data, though its properties are poorly constrained.
The high column density produces a featureless, blackbody-like spectrum, resulting in a pseudo-continuum (as seen in Fig.~\ref{fig:slab_models}) likely associated with warm C$_2$H$_2$ gas at a temperature comparable to the dust continuum. 
However, 
the inferred column density and emitting area depend strongly on the assumed fraction of the continuum attributed to this optically thick C$_2$H$_2$ emission. 

\subsection{Extended molecular hydrogen emission}\label{subsec: molecular hydrogen}
\begin{figure*}
    \centering
    \includegraphics[width=1\linewidth]{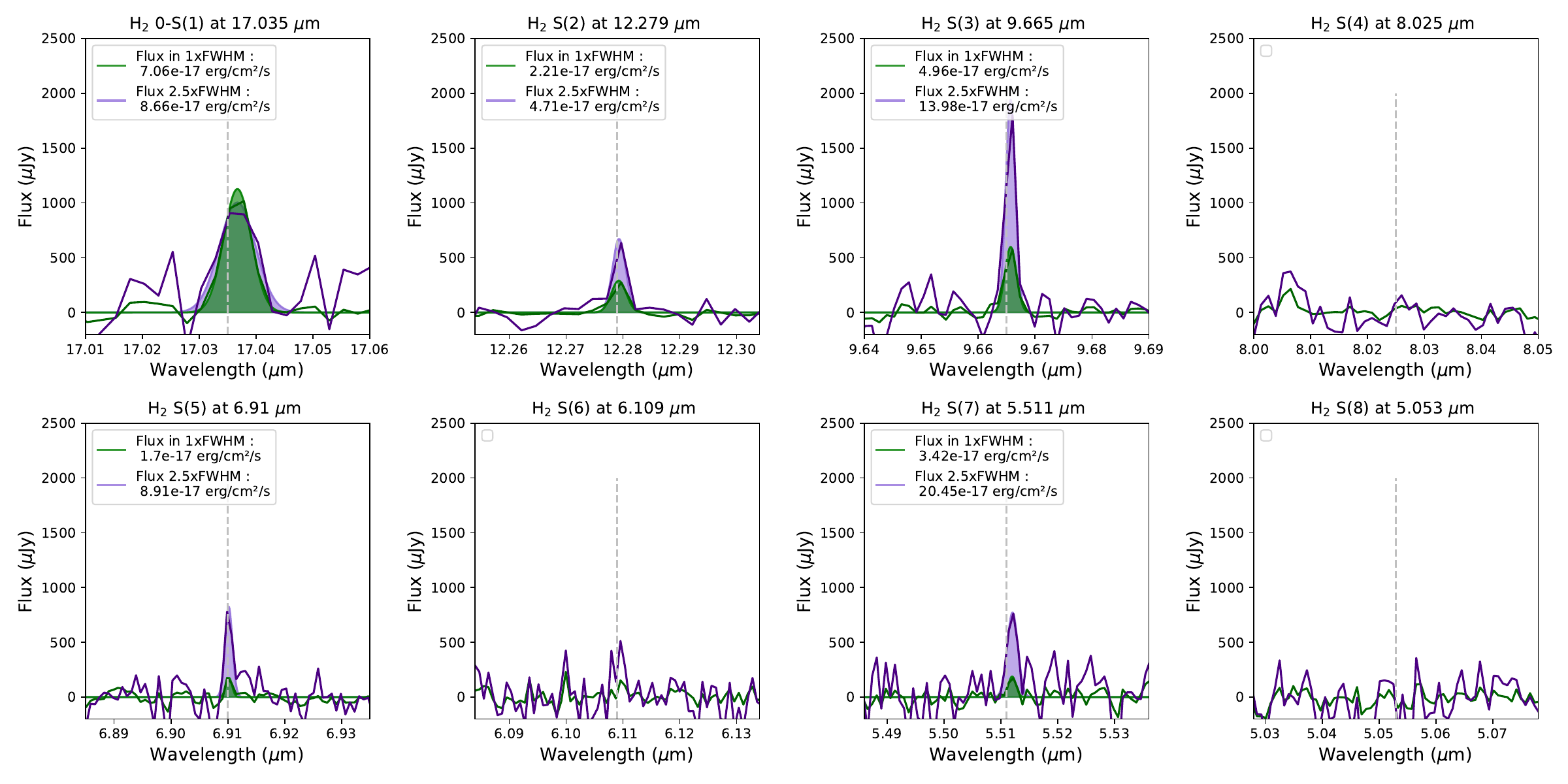}
    \caption{H$_2$ lines in Delorme\,1\,AB\,b spectrum for the two aperture sizes, with the corresponding Gaussian fit also shown.
    The data are at their native resolution, and the fitted model is oversampled.}
    \label{fig:H2_lines_fit_apertures}
\end{figure*}

The observed spectrum also exhibits molecular hydrogen emission lines.
These represent the first H$_2$ emission lines detected from a circumplanetary disk. 
These transitions have been observed and studied at MIR wavelengths \citep[e.g.][]{pascucci_atomic_2013,
carmona_search_2008, 
roueff_full_2019, 
jennings_ground_1983} 
and more recently with MIRI/MRS data in more luminous, higher mass and younger targets
\citep[e.g., SY\,Cha, J160532][]{schwarz_minds_2025,tabone_rich_2023, franceschi_minds_2024}, but only one other of similar age \citep[J0446B,][]{long_first_2025}.
Within the MRS wavelength range, we have access to molecular hydrogen pure rotational lines $\nu$ = 0-0 from S(1) to S(8). 
The 1-D spectrum of Delorme\,1\,AB\,b clearly exhibits the S(1), S(2), S(3), and S(5) lines, as shown in Fig.~\ref{fig:spectra}, S(5) being only weakly detected.
For each line, we remove the continuum and fit its emission using a Gaussian profile centered at the corresponding wavelength.
To obtain a robust estimate of the uncertainty on the line flux, we add in quadrature the 5\% absolute flux calibration uncertainty \citep{argyriou_jwst_2023} to the root-mean-square noise measured from the spectrum. 
We then generate 1000 synthetic realizations of the spectrum using a Monte Carlo approach, where Gaussian noise consistent with the total uncertainty is added to the observed flux. 
For each realization, we fit again the emission line with a Gaussian profile and integrate the resulting model to measure the line flux. 
The distribution of these Monte Carlo flux measurements is well approximated by a Gaussian, from which we derive the final line flux and its 1$\sigma$ uncertainty.
The resulting fits are displayed in green in Fig.~\ref{fig:H2_lines_fit_apertures} and the corresponding line parameters are summarized in Table \ref{tab:h2_lines}.
This procedure was repeated using fluxes extracted with various aperture sizes.
Specifically, we extracted a spectrum using an aperture of 2.5 FWHM of the PSF, following the method described in Sect. \ref{sec:data_red} and applying the appropriate aperture correction factors.
For comparison, we also extracted a spectrum using a fixed physical aperture size of 40~au.

\begin{table*}[ht]
    \centering
    \caption{H$_2$ spectral line properties and fluxes values extracted with different apertures.}
    \begin{tabular}{lcccccccc}
    \hline
    \hline
    {Line} & {Wavelength} & {$E_{\rm u}$} & {$g_{\rm u}$} & {$A_{\rm ul}$} & \multicolumn{3}{c}{Flux [$10^{-17}$ erg\,cm$^{-2}$\,s$^{-1}$]}\\
           & [$\mu$m]     & [K]     &         & [$10^{-10}$ s$^{-1}$] & 1×FWHM & 2.5×FWHM & 40 au \\
    \hline
    H$_2$ S(1) & 17.04 & 1015.1 & 9  & 2.94 & $7.06 \pm 0.86$ & $8.66 \pm 2.70$ & $5.86 \pm 0.75$ \\
    H$_2$ S(2) & 12.28 & 1681.7 & 5  & 4.76 & $2.21 \pm 0.88$ & $4.71 \pm 1.11$ & $3.13 \pm 0.92$ \\
    H$_2$ S(3) & 9.67  & 2504.1 & 21 & 9.80 & $4.96 \pm 1.22$ & $13.98 \pm 1.39$ & $10.25 \pm 1.20$ \\
    H$_2$ S(4) & 8.03  & 3474.5 & 9  & 14.3 & -- & -- & -- \\
    H$_2$ S(5) & 6.91  & 4586.3 & 33 & 21.4 & $1.70 \pm 2.25$ & $8.91 \pm 2.05$ & $7.49 \pm 1.96$ \\
    H$_2$ S(6) & 6.11  & 5829.8 & 13 & 29.5 & -- & -- & -- \\
    H$_2$ S(7) & 5.51  & 7195.9 & 45 & 39.1 & $3.42 \pm 9.50$ & $20.45 \pm 6.67$ & $17.62 \pm 6.69$ \\
    H$_2$ S(8) & 5.05  & 8683.8 & 17 & 49.9 & -- & -- & -- \\
    \hline
    \end{tabular}
    \tablefoot{Fluxes are given for three apertures: 1×FWHM (used in Figure~\ref{fig:spectra}), 2.5×FWHM, and a fixed 40~au physical aperture
    (the 40~au aperture does not reflect the absolute flux, as no aperture correction has been applied due to the varying PSF sizes in this aperture across wavelengths).}
    \label{tab:h2_lines}
\end{table*}

Moreover, comparing with the CPD slab models (Sect. \ref{sec:cpd_molec}), we confirmed that none of the H$_2$ emission lines overlap with detected hydrocarbon emission lines.
Although C$_2$H$_2$ exhibits spectral features near the H$_2$ S(2) line, the H$_2$ emission is considerably stronger than that of the hydrocarbon.

The lines appear slightly redshifted, with the peak of the Gaussian fit showing offsets ranging from 5 to 30 km/s.
This could be consistent with outflows; however, such shifts might also arise from line undersampling and wavelengths calibrations.

\begin{figure*}
    \centering
    \includegraphics[width=1\linewidth]{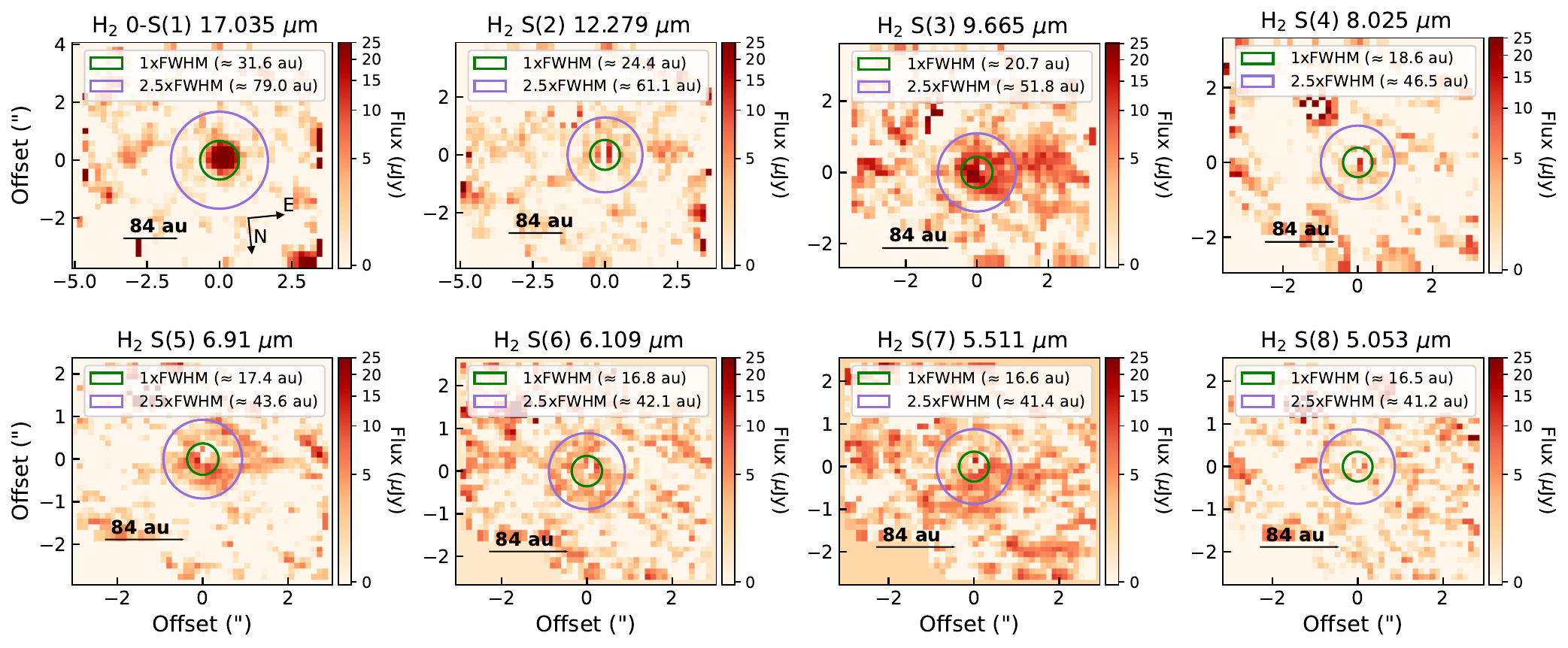}
    \caption{Residuals after PSF subtraction for each H$_2$ line. Circles indicate concentric apertures at 1\,$\times$ FWHM in green and 2.5\,$\times$ FWHM in purple. 
    This physical scale in au depends on the pixel scale, which varies across spectral bands.}
    \label{fig:Residuals_H2lines}
\end{figure*}

We perform a PSF-subtraction of the cubes to determine if this emission is extended and spatially resolved, in order to understand its origin.
First, we sum the cube using the four slices surrounding each H$_2$ line to create the H$_2$ image.
Then, we construct a PSF image by using the 30 neighboring slices on each side of the H$_2$ line, but 5 slices away from the H$_2$ line wavelengths. 
Finally, we subtract these PSF images from the H$_2$ image to obtain a residual image \citep{worthen_miri_2024, pontoppidan_high-contrast_2024, arulanantham_jwst_2024}.
This follows a similar approach to the analysis presented in \citet{bajaj_jwst_2024, schwarz_minds_2025}, although we did not use any deconvolution algorithm to sharpen the image.
These residuals after the PSF subtraction are shown in Figure~\ref{fig:Residuals_H2lines}, revealing extended emission for some H$_2$ lines.
The S(1) and S(2) lines are not extended, with the emission matching the exact size of the PSF.
This suggests it traces unresolved gas in the disk, likely corresponding to the surface of the CPD. 
In contrast, the S(3) and S(5) lines show clear extended emission, reaching up to 40~au.
The S(4), S(6), and S(8) lines are not detected, either in the extended emission maps or in the extracted spectra. In contrast, the S(7) line becomes marginally detectable when using the largest extraction aperture.
The planetary atmosphere dominates at the wavelengths of the S(6) to S(8) lines, which is one of the factors that likely prevents their detection.
For most lines - except S(1) - strong residuals remain at the host position, likely caused by imperfect PSF subtraction.

\begin{figure*}
    \centering
    \includegraphics[width=1\linewidth]{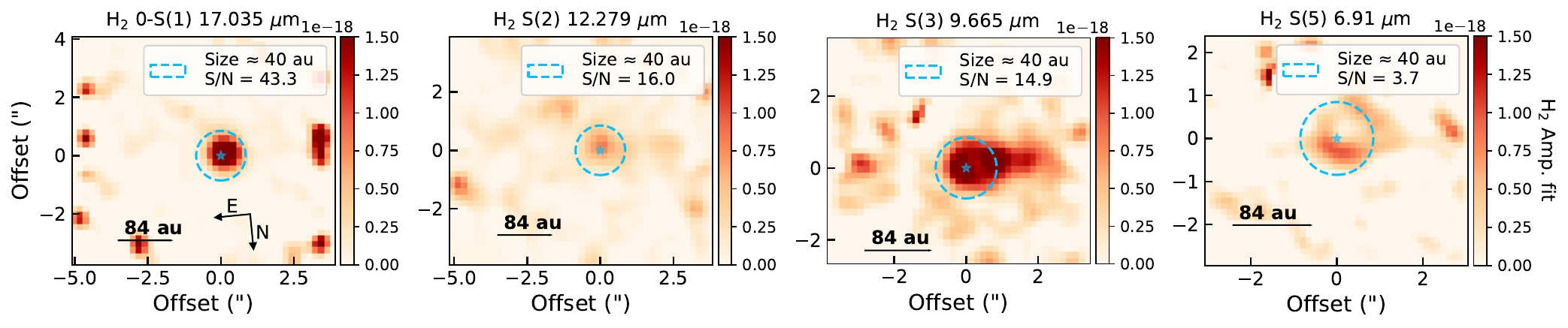}
    \caption{H$_2$ amplitude maps. Each pixel value corresponds to the amplitude of the H$_2$ line Gaussian fit expressed in ergs per square centimeter per second. 
    The dashed circles indicate fixed-radius apertures of 40\,au for comparison. A Gaussian smoothing with $\sigma = 1$ pixel has been applied to each image. 
    The color scale is the same for each image.}
    \label{fig:Residuals_amp_maps}
\end{figure*}

To evaluate the significance of the detection, we generate a map of the amplitude of the fitted H$_2$ line.
For each spaxel, we remove the continuum and fit the H$_2$ line using a Gaussian profile, with the wavelength and standard deviation fixed based on the previous fit. 
As a result, only the amplitude varies in the fit, and this value is used for each spaxel to create the maps shown in Figure~\ref{fig:Residuals_amp_maps}. 
These maps illustrate the amplitude of the Gaussian fit to the H$_2$ line for each pixel, emphasizing both the H$_2$ emission and its spatial extent and allowing us to measure the $S/N$. 
First, we co-added the pixels in an aperture of 40~au, globally corresponding to its spatially extended emission.
The noise $\sigma$ is estimated in the residual maps as the standard deviation of the pixels inside a ring of 10 pixels centered on the planet.
The host binary is excluded from the noise measurement using a circular mask with a radius of 3 pixels, as it exhibits strong residuals due to PSF subtraction.
Therefore, the $S/N$ of the emission is $S/N = F_{40~\mathrm{au}\,\mathrm{aperture}} / (\sqrt N_{\rm px}\times \sigma)$.
This provides $S/N$ ranging from 
$S/N = 3.7$ for the line S(5) to $S/N = 14.9$ for the S(3).
It confirms that the S(4) transition and lines at higher energy are not robustly detected ($S/N$ < 3).
At longer wavelengths, the S(1) and S(2) lines reach $S/N$ values of 43.3 and 16.0, respectively. 
These two lines remain unresolved spatially; although the integrated flux of the S(2) line increases with larger apertures, the gain is insufficient to provide robust constraints on a potential spatial extension. 
Based on the PSF size at a distance of 47.2~pc, this sets an upper limit of 9.44~au for the emitting region at these wavelengths. 
In contrast, the S(3) and S(5) lines are spatially resolved and trace extended emission over a region of approximately 40~au, predominantly oriented toward the north-west.

\subsection{Estimation of the H$_2$ gas temperature}
The measured emission line fluxes can be used to derive the total column density of molecular H$_2$ gas, as well as its excitation temperature, by fitting a line to the rotation diagram \citep{goldsmith_population_1999}.
For a molecular population in LTE, the number of molecules in each upper energy level follows a Boltzmann distribution,
\begin{equation}
\label{eq:Nu_boltzmann}
    N_{\rm u} = \frac{N_{\rm tot}}{Z(T)} g_{\rm u} e^{- E_{\rm u }/ (k_{\rm B} T)},
\end{equation}
where $N_u$ is the population of the upper energy level, $N_{\mathrm{tot}}$ is the total column density, and $Z(T)$ is the partition function, which characterizes the temperature-dependent statistical distribution of states in a system at equilibrium. 
The term $g_{\rm u}$ denotes the statistical weight of the level, and $E_{\rm u}$ is the energy of the upper state. 
These two terms are provided in Table~\ref{tab:h2_lines} for each H$_2$ transition \citep{roueff_full_2019}.
Assuming optically thin emission, the column density can be related to the line flux $F_{\mathrm{line}}$ as follows,
\begin{equation} 
    N_{u} = \frac{4 \pi F_{\mathrm{line}}}{A_u h \nu \Omega}, 
\end{equation} 
where $A_u$ is the Einstein coefficient for spontaneous emission, $h$ is Planck’s constant, $\nu$ is the frequency of the transition, and $\Omega$ is the emitting area, given by $\Omega = \pi (r_{\mathrm{em}}/d)^2$, where $r_{\mathrm{em}}$ is the radius of the emission region. 
We measured the emission line flux 
$\mathit{F}_{\mathrm{line}}$ for each H$_2$ transition in the spectrum and repeated this procedure for the multiple aperture sizes.

\begin{figure}
    \centering
    \includegraphics[width=1\linewidth]{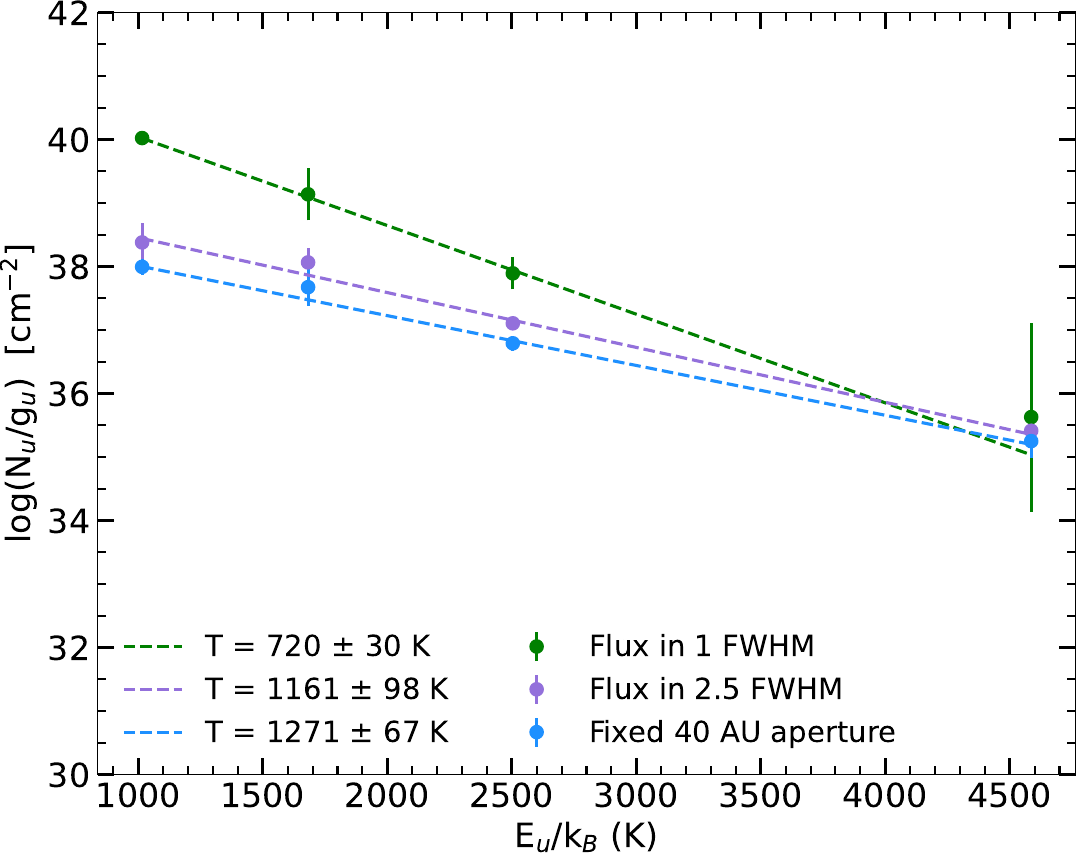}
    \caption{Rotation diagram to measure the excitation temperature of H$_2$.}
    \label{fig:rotation_diagram_1temp}
\end{figure}

The rotation diagram, shown in Fig.~\ref{fig:rotation_diagram_1temp}, is constructed by performing a linear regression between $\ln(N_{\rm u} / g_{\rm u})$ and $E_{\rm u}$. The resulting slope of this regression is given by $-1/(k_{\rm B} T)$, where $T$ is the excitation temperature of the molecular H$_2$ gas around the planet. This relationship is described by the following equation:
\begin{equation}
    \ln(N_{\rm u} / g_{\rm u}) = -E_{\rm u} / (k_{\rm B} T_{\mathrm{exc}}). 
\end{equation} 
We note that this method offers a robust approach to estimating the excitation temperature in systems where the molecular level populations are in LTE and optically thin; as discussed in \citet{franceschi_minds_2024} for example.
Any temperature gradient across the emitting region or non-LTE effects (or opacity effects) will appear as a deviation from the best-fit rotation diagram.  
Given the presence of both extended and compact emission in the line images (see Figure \ref{fig:Residuals_H2lines}), we adopt various aperture size.
We compare the temperature measurement with flux derived from 1$\times$ and 2.5$\times$ FWHM apertures and we also used a fixed physical aperture size of 40\,au across all transitions to ensure consistent spatial sampling of the emitting region for flux measurement.
For each aperture, we fit the rotational diagram using a single-temperature component.  
Uncertainties are derived by propagating the flux measurement errors, accounting for their contribution to the covariance in the fit.   
We find that the derived excitation temperature increases with increasing aperture size.
Moreover, we note that the H$_2$ emission, comprising both extended and compact components, may trace different physical regions.  
In principle, the rotational diagram could be fitted with multiple temperature components; however, this is not reliable given that only four lines are robustly detected.
The rotation diagram provides a measurement of 
T = 1161$\pm$98\,K for the extended component, whereas the non-extended component exhibits a lower temperature of T = 720$\pm$30\,K.
For comparison, the fixed 40\,au aperture in physical size yields a slope of T = 1271$\pm$67\,K, close to the value measured for the extended component.
We excluded the S(7) line from this measurement due to its faintness, which could bias the results.
However, we note that including the S(7) line in the analysis results in a significantly higher excitation temperature, with an upper limit of $T < 1700\,K$, as indicated by the slope of the rotation diagram, although this leads to a poorer fit.
In Sect.~\ref{sec:wind}, we consider whether the extended component of the H$_2$ emission may trace an outflow, such as a disk wind, potentially contributing to the dispersal of the CPD.\\

We compute N$_{tot}$ using equation \ref{eq:Nu_boltzmann}, for which we use the partition function for H$_2$ following \citep{popovas_partition_2016} treating ortho- and para-H$_2$ separately: 
$Z_{\rm para} = g_{\rm J} e^{-E_{\rm J}/T}$ and $Z_{\rm ortho} = 3g_{\rm J} e^{-E_{\rm J}/T}$, with $g_J = 2J+1$ and $E_J = BJ(J+1)$. 
Assuming an ortho-to-para ratio (OPR) = 3, we weighted the partition function with: Z(T) = [OPR/(1+OPR)] $\times$  $Z_{\rm ortho}$ +  [1/(1+OPR)] $\times$  $Z_{\rm para}$.
Using the rotation diagram, we can solve for $N_{\rm tot} = Z(T) e^{\rm b}$, with b the intercept of the diagram, and then estimate the mass of warm molecular hydrogen using the relation
\begin{equation}
    M_{\text{H}_2} = 2\,m_{\text{H}}\, N_{\text{tot}}\, A,
\end{equation}
where $m_{\text{H}}$ is the atomic mass of hydrogen, and $A = \Omega \times d^2$ is the projected physical area of the emitting region.
We adopt a distance $d = 47.2\,\text{pc}$, and compute $A$ using the various aperture sizes.
The molecular mass of $H_2$ is $m_{\text{H}_2} = 3.347 \times 10^{-24}\,\text{g}$.
These values are summarized in Table~\ref{tab:h2_prop}.
\begin{table*}[ht]
\centering
\caption{Properties of H$_2$ with the two aperture sizes.}
\begin{tabular}{c|ccccc}
\hline
\hline
Apertures & T$_{\mathrm{exc}}$ & Z(T) & N(H$_2$)  & Mass H$_2$  & $\dot{M}$(H$_2$) \\
 & (K) & & [$\times 10^{18}$ cm$^{-2}$] & [$\times 10^{-6}$ M$_{\mathrm{Jup}}$] & [$\times 10^{-10}$ M$_{\mathrm{Jup}}$/yr] \\
\hline
1×FWHM     & $720 \pm 30$   & 10.97 & $10.8 \pm 0.7$  & $1.02 \pm 0.06$ & $1.79 \pm 0.11$ \\
2.5×FWHM   & $1160 \pm 100$ & 17.42 & $2.05 \pm 0.27$ & $1.21 \pm 0.16$ & $2.12 \pm 0.28$ \\
\hline
\end{tabular}
\tablefoot{The mass-loss rate calculation are presented in Sect. \ref{sec:wind} (eq. \ref{eq:mass_loss_rate}). It assumes a launched radius of 40\,au.}
\label{tab:h2_prop}
\end{table*}
The mass of warm H$_2$ is about 
$(1.2 \pm 0.2) \times 10^{-6} \, M_\mathrm{Jup}$
(which is about twice the mass of Ceres, or approximately 3\% of the Moon's mass).
Finally, we note that this H$_2$ emission traces the hot/warm inner CPD gas in the inner disk, which may not be representative of the bulk gas reservoir in the disk that is likely at lower temperature; and which may be better observed at longer wavelengths (e.g. with ALMA).

\section{Discussion}
\label{sec:discussion}

\subsection{Atmospheric characterization and the influence of the CPD emission}
The inferred properties of the planet are characteristic of young, directly imaged exoplanets, with an averaged temperature of $T_{eff} = 1725 \pm 134$K.
Among the four atmospheric models, \texttt{ATMO} and \texttt{Exo-REM} yield slightly better $\chi^{2}$ values compared to the \texttt{Sonora} and \texttt{BT-Settl} models.
The retrieved atmospheric composition indicates various C/O ratio and metallicity depending on the adopted atmospheric models.
For example, the values for \texttt{Exo-REM} are extremely supersolar in [Fe/H], and the \texttt{ATMO} models are extremely subsolar.
Degeneracies between inferred log\,\textit{g}, metallicity, and cloud properties can complicate abundance estimations for giant planet atmospheres \citep{molliere_retrieving_2020}. 
If the planet's C/O ratio is indeed significantly enhanced (e.g.,\texttt{ATMO} and \texttt{Exo-REM} best-fits including a blackbody), 
such a composition could be consistent with formation between the water and CO snowlines, where the gas phase becomes carbon-rich due to oxygen depletion, and the solid phase is enriched in ices.
Comparing the planetary abundance with the composition of the host star can trace potential formation location within a circumstellar disk and subsequent dynamical history of the planet \citep{oberg_effects_2011}, as a super-stellar metallicity would point to migration involving the significant accretion of solids. 
However, the stellar metallicity and C/O ratios for the host binary have not yet been measured.
The need to jointly consider stellar composition, atmospheric abundances, migration scenarios, and CPD properties underscores the complexity in detailed modeling of these systems. 

Nonetheless, recent discoveries of low-mass M-type stars hosting carbon-rich disks \citep{pascucci_atomic_2013, tabone_rich_2023, arabhavi_abundant_2024} offers an intriguing potential explanation for elevated C/O for the wide-orbit planet, if the companion formed near the host binary and later migrated outward.
Indeed, most directly imaged planets show elevated C/O ratios \citep{hoch_assessing_2023} with Delorme\,1\,AB\,b exhibiting one of the highest measured values.
However, these estimates are strongly model-dependent and generally remain poorly constrained for the majority of giant exoplanets.
The hydrocarbon-rich nature of the CPD similarly lends credence to this hypothesis, although the present wide separation and large planet mass relative to the mass of the host stars remain difficult to reconcile with standard core accretion models and measured disk masses of low-mass stars.
The elevated C/O ratio further supports the scenario of a disk that becomes carbon-rich early in its evolution, as expected around low-mass M-type stars \citep{mah_close-ice_2023}.
Moreover, \citet{pegues_atacama_2021} found that the outer disks around M dwarfs also tend to be carbon-rich, supporting this hypothesis.

Given the high accretion rate, it is likely that the planetary atmosphere is being replenished by the CPD.
The C-rich nature of the CPD could therefore also explain the potentially elevated C/O ratio observed in the planetary atmosphere. 
Since accretion is known to be highly variable for this target \citep{betti_near-infrared_2022, demars_exoplanet_2025}, it may have a significant impact on both the atmospheric composition and the evolving properties of the CPD. 
Multi-epoch observations that simultaneously probe the atmosphere, accretion activity and CPD composition would offer valuable insights into the interplay between these components.

The wide wavelength range and the overlapping contributions from the planet's atmosphere and CPD molecular features make it challenging to accurately constrain all atmospheric parameters using a simple forward modeling approach.
A full retrieval analysis of the spectrum over the entire wavelength range is beyond the scope of this work but would offer a complementary study of the planet's atmosphere to provide precise atmospheric parameters, and derive abundances of molecules to further understand its formation pathway \citep[e.g.,][]{nasedkin_four---kind_2024}.
The range from 8 to 12\,$\mu$m is then likely a mixture of planetary atmosphere components and circumplanetary disk components, containing both molecular gas and possibly dust and clouds in the atmosphere.
A joint modeling approach using planetary atmospheric models and CPD models would be relevant to model this part of the wavelength range, which we leave for future work.
Integrating slab models for the CPD into an atmospheric forward-modeling framework would be a valuable next step for characterizing such objects. 
While this approach may be computationally intensive because of significant parameter degeneracies, the use of strong priors (such as those derived from detailed disk chemistry models) can help to mitigate these difficulties and improve parameter constraints.

Similarly, coupling atmospheric models with evolutionary models can improve parameter constraints, provided the system’s age is reasonably well known. 
A practical starting point would be to apply robust atmospheric priors from NIR data and use slab models in the MIR, fitting both only in overlapping regions to reduce model complexity.
The question of at which wavelength the CPD features begins to appear in the spectrum remain uncertain, and veiling from accretion or from the CPD itself may contribute to the observed spectral features.
Additionally, it would be interesting to consider whether veiling from CO in emission within the CPD could be present, and whether the atmospheric models predict more CO at this temperature (as shown in  Fig\,\ref{fig:Spectrum_H2O_CO} which display the molecular templates spectrum given the best-fit atmospheric models).
To properly constrain CO, coverage in the NIR range is essential, as the MIR only captures the tail end of the CO fundamental ro-vibrational band, for example with MIRI observations \citep{francis_joys_2024, grant_full_2024,temmink_minds_2024, salyk_spitzer_2011}, but also with ground-based instruments \citep{brown_vlt-crires_2013, banzatti_scanning_2022}.
This would also enable simultaneous modeling of both CO components: absorption from the planetary atmosphere and emission from the CPD.
Distinguishing the CPD and the planet's atmosphere would be possible at higher spectral resolution, as the circumplanetary gas is expected to exhibit a different velocity than the planet's atmosphere. 
Currently, no high-resolution spectrograph is available at MIR wavelengths. 
Although CRIRES+ and, in the near future, ELT/METIS provide high-resolution integral-field spectroscopy in the L and M bands (R$_\lambda \sim 100{,}000$), the circumplanetary disk around this planet is likely too faint between 3 and 5$\mu$m to be detectable.
Nevertheless, these instruments could help probe the CPD's indirect influence; for example, through changes in the shape or intensity of CO features in the planetary spectrum.

\subsection{Circumplanetary disk structure}
\paragraph{Dust cavity.}
Crystalline and amorphous dust features are commonly observed in MIR disk spectra, but this MIRI/MRS spectrum does not show any silicate features around 9 – 11\,$\mu$m, suggesting an absence of small silicate grains.
This contrasts with the YSES-1\,b exoplanet, which shows evidence of a circumplanetary disk containing micron-sized silicate grains, as revealed by MIRI low-resolution data \citep{hoch_silicate_2025}.
The lack of the 10\,$\mu$m silicate feature indicates that either the dust grains have grown and settled out of the upper layers of the CPD into the optically thick disk midplane \citep{kesslersilacci_c2d_2006,kessler-silacci_probing_2007}, or that the inner part of the CPD is depleted of silicate dust.
If the dust population is dominated by larger grains, the sizes would need to exceed 5\,$\mu$m, such that they could no longer produce distinctive spectral features in this wavelength range 
\citep[cf.][]{tabone_rich_2023, arabhavi_abundant_2024}.
This implies that the dust population is dominated by larger grains, potentially with sizes exceeding 5\,$\mu$m, that no longer produce distinctive spectral features in this wavelength range.

We computed the dust sublimation radius, defined as the distance from the planet at which the CPD temperature reaches 1500\,K and interior to which dust is expected to sublimate \citep{dullemond_inner_2010} using
\begin{equation}
  R_{\mathrm{sub}} \approx \left( \frac{L_{\mathrm{pl}}}{16\pi \sigma T_{\mathrm{sub}}^4} \right)^{1/2}.
\end{equation}
The sublimation radius is \( R_{\mathrm{sub}} = 6 \times 10^{-4} \,\mathrm{au} \) (i.e., \(1.2\,R_{\mathrm{Jup}}\)), using $L_{\mathrm{pl}} = 10^{-3.5} L_{\odot}$ from the atmospheric fit values.
It corresponds to the dust evaporation boundary: dust can exist beyond this radius but disappears closer in due to the higher temperatures.

From the blackbody modeling of the CPD that is tracing the dust, we can estimate the radius of the cavity in the CPD. 
The radiation from the stellar binary is negligible at this separation (the flux is reduced by r$^{2}$, i.e., the radiation flux at a distance of 84\,au is reduced by a factor of 7056 compared to the flux measured at 1\,au from the star).
Using Stefan–Boltzmann’s law and the parameters derived in the atmospheric modeling (Sect. \ref{sec:atm_charact}), we measured: 
\begin{equation}
    R_{cav}^{BB} = \sqrt{\frac{L_{\rm B}}{16 \pi \sigma T^{4}_{BB}}}
\end{equation}
with $T_{BB}$ = 295$\pm$27\,K, 
and $L_{B}$ = 10$^{-3.51\pm 0.02} L_{\odot}$.
This indicates a cavity of 
$R_{\rm cav}^{BB} = 32.6 \pm 3.1 R_\mathrm{Jup}$ (0.016~au),
as done in \cite{cugno_mid-infrared_2024}, in the case of GQ\,Lup\,b, for which they measured a cavity of $40.3\,R_\mathrm{Jup}$.
This likely leads to an overestimation of the cavity size, since if some of the planet’s emission is scattered by the CPD material, the actual cavity would be smaller. This calculation assumes that dust grains do not reflect any radiation coming from the planet.
Since our observations do not detect the warmer dust component, the dust distribution does not appear to extend down to the sublimation radius.
The dust emission traces a colder component ($\sim$300\,K) than the sublimation temperature, suggesting the presence of a dust cavity, since the dust is cooler than expected if the disk extended all the way down to the planetary surface.
The cavity could have several possible origins, among which moon formation could be one of the hypotheses.\\

\noindent
\paragraph{Outer radius.}
Inside the R$_\mathrm{Hill}$ the planet dominates gravitational interactions, this corresponds to the separation until which a CPD can exist.
Simulations and analytic work show that the disk does not completely fill the Hill radius, and that the boundary of the disk is at R$_\mathrm{out}$ = 0.4 R$_\mathrm{Hill}$ \citep[this value can depend on the disk viscosity and its vertical structure][]{martin_tidal_2011, adams_general_2025}.
Using a planet of 14\,M$_\mathrm{Jup}$ at 84~au for a 0.36 M$_{\odot}$ system \citep{delorme_direct-imaging_2013}, the R$_\mathrm{Hill}$ is 19.43~au,
corresponding to a truncation radius of R$_\mathrm{out}$ = 7.8~au ($1.6 \times 10^{4}\,R_{\mathrm{Jup}}$).
The extended emission observed with the H$_2$ lines extends beyond the outer radius of the CPD, supporting its interpretation as a outflowing gas originating from the circumplanetary environment.\\

\noindent
\paragraph{Temperature and density distribution.}
We initially use a single-temperature blackbody model.
Each molecular component of the CPD is also modeled assuming a uniform temperature and density, except for C$_2$H$_2$, which is represented by two distinct reservoirs, each assuming a uniform temperature and density.
Although this provides a reasonably good first-order approximation, the physical reality is likely more complex, involving gradients in both temperature and density. 
For instance, in the case of C$_2$H$_2$, the gas may be hidden in the optically thick mid-plane of the disk, 
with a vertical gradient leading to increasingly diffuse gas in the upper layers, eventually reaching an optically thin regime as seen in T Tauri disks \citep{arulanantham_jdisc_2025}.
Unlike typical T Tauri disks, where the dust continuum can be modeled by a series of blackbodies, the continuum here is reproduced with a single blackbody temperature.
MIRI observations primarily detect only the warmer, upper regions of the disk, since cold, dense C$_2$H$_2$ deeper in the disk are difficult to observe.
Various disk models, such as thinner structures, inclusion of accretion 
\citep{woitke_2d_2024}
or puffed-up inner rims, could be more appropriate than using a blackbody estimation.
As a conclusion, the mid-infrared spectrum provides a unique insight into the chemical composition and structure of the disk.\\

\noindent
\paragraph{Accretion and snowlines.}
High accretion rates are also expected to influence the structure of the disk.
For instance, an accretion outburst can push the snowline outward, altering the temperature gradient within the disk \citep[e.g. for Tauri stars][]{houge_collisional_2023, smith_jwsts_2025, kospal_time-resolved_2025}.
However, the extent and impact of such processes remain an open avenue of investigation for CPDs.
A variation of 100\% in line luminosity \citep{demars_exoplanet_2025} leads to a 44\% shift in the location of the water snowline, based on the relationship from \cite{mulders_snow_2015}.
This equation has been used by \cite{grant_minds_2024} to demonstrate the impact of accretion on the snowline in the DF Tau system. 
It assumes a steady-state viscous protoplanetary disk that is optically thick to its own radiation. However, this CPD is likely time-variable due to its highly variable accretion, and the optically thick assumption may not hold across the entire disk.
We assume a mean accretion rate of $2\times10^{-9}\, M_\mathrm{Jup}/\text{yr}$ (averaged over two epochs from \cite{betti_near-infrared_2022} and derived using the stellar scaling of line luminosity to accretion luminosity).
The snowline in the CPD is located at 0.012~au from the companion Delorme 1 AB b.
However, using a higher accretion rate of $5\times10^{-8}\, M_\mathrm{Jup}/\text{yr}$ (derived via planetary scaling of line luminosity to accretion luminosity), it moves outward to 0.045~au.
These different values highlight that the snowline position depends not only on the assumed method for estimating accretion but also on the intrinsic variability of the accretion itself.
The enhanced UV radiation from accretion hot spots raises the gas temperature, thereby strengthening molecular emission lines.
At the same time, increased UV flux can lead to greater photodissociation of certain species, altering molecular abundance ratios.
These effects have been demonstrated for T Tauri stars \citep{najita_fuv_2017} and
these observational data open avenues for disk thermochemical models.
While previous CPD models have primarily focused on CO millimeter lines \citep{rab_chemistry_2017}, which trace a different gas reservoir, the impact of accretion on the disk chemistry remains largely unexplored, highlighting exciting directions for future studies.
Disk chemistry is therefore likely to be strongly affected by accretion processes, although current constraints are limited.
Multi-epoch observations of this disk, along with measurements of accretion rates at the same epoch,
could assess how accretion affects disk chemistry, carbon content, and potentially moon formation in the CPD.\\

\begin{figure}
    \centering
    \includegraphics[width=1\linewidth]{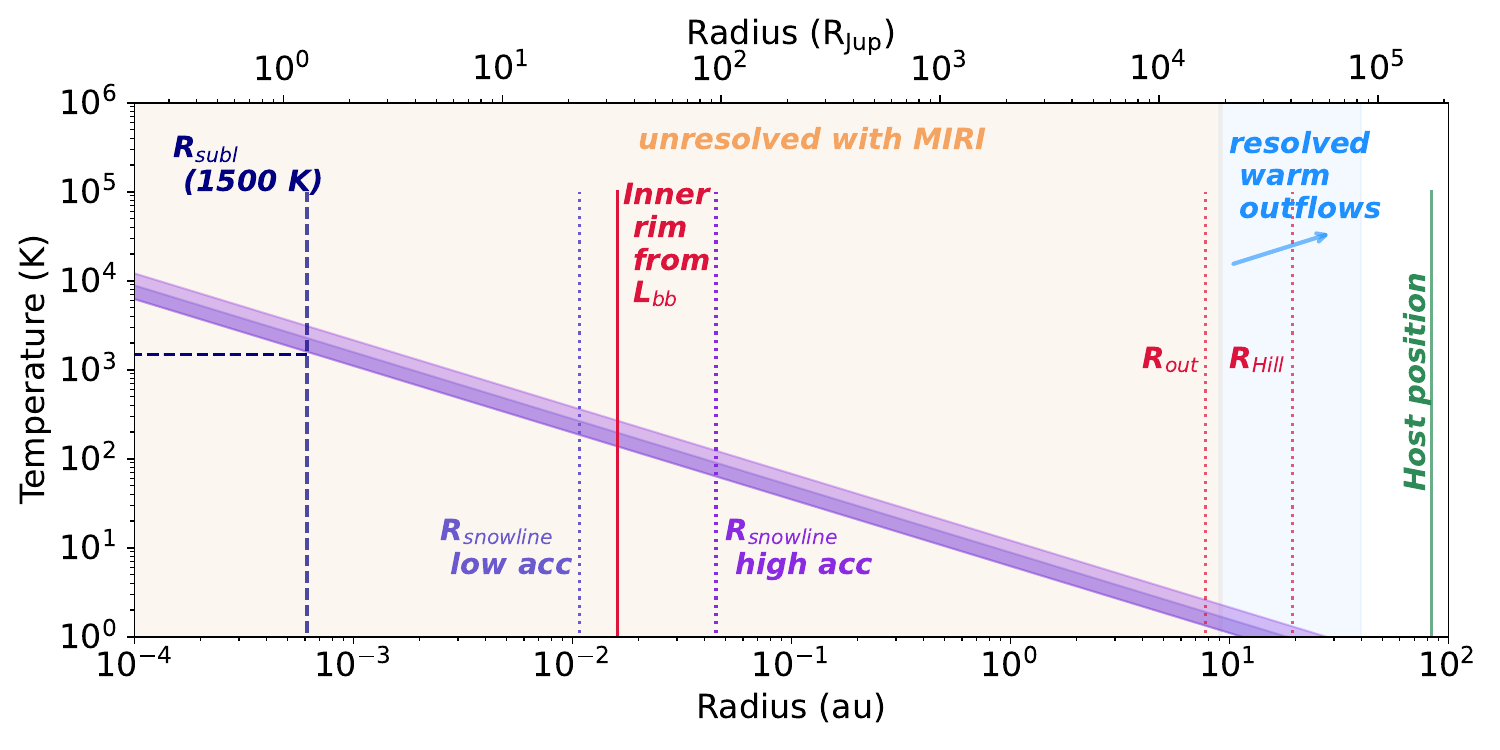}
    \includegraphics[width=1\linewidth]{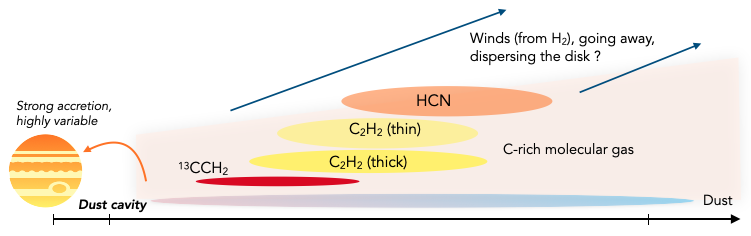}
    \caption{Schematic of the Delorme\,1\,AB\,b system: the planet and its CPD.
    Top: Temperature as a function of radius with both axes on a logarithmic scale.
    The upper and lower purple bounds correspond to scenarios with high and low accretion, respectively.
    Representative radius values are indicated. 
    The beige region denotes scales below the spatial resolution of MIRI/MRS.
    Bottom: Illustration of the CPD structure with the molecular gas species, intended as a simplified representation; it does not necessarily reflect the true spatial distribution of molecular species.}
    \label{fig:structure}
\end{figure}

\noindent
\paragraph{Structure overview.}
Figure~\ref{fig:structure} summarizes the radial scales discussed in this section. The sublimation radius R$_\mathrm{subl}$ is estimated assuming a dust temperature of 1500\,K and lies interior to the inner rim inferred from a blackbody fit to the circumplanetary dust emission at 300\,K. 
This suggests the likely presence of a dust cavity.
The C-rich molecular gas components remain spatially unresolved with MIRI/MRS; their locations in the schematic (bottom plot) are inferred based on assumed density and vertical temperature structures.
We note that the radii derived from slab model fitting correspond to effective emitting areas and do not reflect precise spatial positions.
At larger separations, we indicate the Hill radius and the corresponding truncation radius, defined as R$\mathrm{out}$ = 0.4\,R$_\mathrm{Hill}$.
The H$_2$ emission lines extend up to twice the Hill radius and are spatially resolved with MIRI. They are located at approximately half the separation from the host binary (84\,au).
Finally, the horizontal temperature structure is shown: the temperature as a function of radius is plotted shaded purple areas, following the prescription of \citet{adams_infrared_1986}. 
The upper temperature limit corresponds to the case with the highest accretion rates, while the lower limit represents the case without accretion, assuming mass accretion rates of $2 \times 10^{-9}$ and $5 \times10^{-8}\,M_\mathrm{Jup}\,\mathrm{yr}^{-1}$, respectively.
The values of the snowline radii corresponding to these two values are also indicated.

\subsection{Disk chemical evolution around low-mass objects}

The spectrum of the circumbinary companion  Delorme\,1\,AB\,b resembles those of inner disks around more massive and isolated objects, such as BDs, VLMS (<0.3 M$\odot$), but also young T Tauri stars.
Most of these objects, recently observed with MIR spectroscopy, are younger than the Delorme\,1 system, except for the disk around the J0446B, which is of comparable age ($\sim$30–45 Myr; \citealt{long_first_2025}).

Similar to these BD and VLMS disks, the CPD of Delorme\,1\,AB\,b shows consistent detections of HCN and C$_2$H$_2$ \citep{arabhavi_minds2_2025}, although it exhibits lower and cooler dust content and a more limited set of detectable hydrocarbons — even when compared to J0446B.
Overall, Delorme\,1\,AB\,b exhibits a more limited hydrocarbon inventory than BD and VLMS disks, with no detections of other species such as C$_4$H$_2$ and C$_6$H$_6$, which are now commonly observed with MIRI/MRS in isolated BDs and VLMSs \citep[e.g.,][]{tabone_rich_2023}.
Given the faintness of this CPD relative to VLMS and BD disks, the line-to-continuum sensitivity is reduced.
However, this alone cannot account for the more limited molecular inventory observed.
For instance, TWA\,27\,A (also known as 2M\,1207\,A),  which shows a similarly low continuum flux level \citep{patapis_jwstmiri_2025}, exhibits a significantly richer hydrocarbon spectrum.
This comparison shows that the simpler chemical composition seen in the CPD around Delorme\,1\,AB\,b is not only a consequence of lower $S/N$ on the continuum, but rather reflects intrinsic differences in the disk chemistry.

More surprisingly, no oxygen-bearing molecules are detected in the CPD around Delorme\,1\,AB\,b, with the absence of H$_2$O emission consistent with findings by \citet{arabhavi_minds_2025}, who report that C$_2$H$_2$ emission dominates over water in disks around VLMSs and BDs compared to those around T Tauri stars.
This contrasts especially with young T Tauri stars, which typically exhibit water-rich spectra \citep[e.g.][]{carr_organic_2008, carr_organic_2011, pontoppidan_spitzer_2010, salyk_spitzer_2011, salyk_high-resolution_2019, banzatti_depletion_2017, banzatti_jwst_2023, banzatti_water_2025}.

The CPD of Delorme\,1\,AB\,b appears depleted in oxygen-bearing species, with no detected H$_2$O or CO$_2$ features.
While the absence of CO$_2$ is not unusual compared to Class II T Tauri disks \citep{salyk_spitzer_2011, arulanantham_jdisc_2025},  it stands in contrast to VLMS and BD disks, where CO$_2$ is consistently observed \citep{arabhavi_minds2_2025}.
However, we note that a weak H$_2$O signal is detected via cross correlation in band 4A ($S/N = 4$), but it is too faint to constrain any water properties (see Appendix~\ref{sec:molecular_mapping}).

A clear transition in disk composition and chemical evolution is observed between stars with >0.2\,M$\odot$ and <0.2\,M$\odot$, marking the divide between solar-type stars and VLMS \citep{grant_transition_2025}. 
This particular object falls within the continuity of the < 0.2\,M$\odot$ group, suggesting a continuity of chemical properties down to planetary-mass objects, though a larger sample will be needed to confirm this trend. Additional JWST/MIRI observations of CPDs are already scheduled \citep[PID 7538, PID 6086,][]{cugno_giants_2025, ward-duong_first_2024}.

We report the first detection of molecular line emission from CPD suggesting that CPDs may have similar chemical characteristics to disks around VLMS.
This observation suggests that, below this mass threshold, disks may follow a common evolutionary path regardless of the central mass.
In this context, Delorme\,1\,AB\,b seems to be evolving as a self-contained system, with disk properties and chemistry developing independently in a manner akin to isolated substellar or planetary systems \citep[e.g.][]{flagg_detection_2025}.

Table~\ref{tab:comparison_targets} provides a comparison with recently observed targets using MIRI/MRS, including very low-mass stars such as J160532 \citep{tabone_rich_2023, franceschi_minds_2024} and J0446B \citep{long_first_2025}, both of which show H$_2$ emission at MIR (extended in the case of SY~Cha; \citealp{schwarz_minds_2025}).
Accretion luminosities were estimated using the mean accretion rates and the method described by \citet{alcala_x-shooter_2014},
\begin{equation}
    L_{\text{acc}} = \frac{G M_{\text{acc}}}{1.25 R_\star}.
\end{equation}
This sample was selected to place Delorme\,1\,AB\,b in context with similar targets or typical properties of a disk around more massive objects, but it is not intended to be comprehensive.
\begin{table*}[t]
    \centering
    \caption{Comparative table of Delorme\,1\,b's disk and disks around more massive objects observed with MIRI/MRS.}
    \begin{tabular}{l|ccc|cc}
       \hline
       \hline
       Properties  & Delorme 1\,AB\,b & J0438 & J0446B & Typical & Typical  \\
       \hline
       Host type & Exoplanet & Brown dwarf  & VLMS & T Tauri &  VLMS/BD\\
       Age (Myrs) & 30 & $\sim$0.6 & 34 & 1 - 10& 1 - 10\\
       Spectral Type & L0 & M7.25 & M4.5  & K0-M5 & M4.5-M9\\
       Distance (pc) & 47.2 & 140.3  & 82 & -- & --\\
       M$_{\star}$ ($M_{\odot}$) & 0.013 & 0.05 & 0.18 & $\sim$ 0.2 - 2 & $\sim$ 0.02 - 0.2\\
       L$_{\star}$ ($L_{\odot}$) & 0.0003 & 0.002 & 0.016 & $\sim$0.01 – 0.2 & $\sim$ 0.1 - 0.002\\
       $\dot{M}$ ($M_{\odot}$/yr) & 0.2–5$\times10^{-11}$ $^{a}$ & 1.58$\times10^{-11}$ & 2.5$\times10^{-11}$ & $10^{-7}$ - $10^{-9}$ & $10^{-9}$ - $10^{-12}$\\
       L$_\mathrm{acc}$ / L$_{\star}$ & 1–37\% & 4\% & 2.5\% & 1-100\% & 0.1–10\% \\
       \hline
       \multicolumn{6}{c}{Molecular emission} \\
       \hline
       H$_2$O & -- & \cmark & \cmark$^{b}$ & \cmark & \cmark$^{b}$ \\
       CO$_2$ & -- & \cmark & \cmark & \cmark & \cmark \\
       OH & -- & -- & -- & \cmark & -- \\
       HCN & \cmark & \cmark & \cmark & \cmark & \cmark\\
       CH$_4$ & -- & -- & \cmark & -- & \cmark \\
       CH$_3$ & -- & -- & \cmark & -- & \cmark \\
       C$_2$H$_2$ & \cmark & \cmark & \cmark & \cmark & \cmark \\
       $^{13}$CCH$_2$ & \cmark & -- & \cmark & -- & \cmark\\
       C$_2$H$_4$ & -- & -- & \cmark & -- & \cmark\\
       C$_3$H$_4$ & -- & -- & \cmark & --  & \cmark\\
       C$_4$H$_2$ & -- & -- & \cmark & -- & \cmark\\
       C$_6$H$_6$ & -- & -- & \cmark & -- & \cmark\\
       C$_2$H$_6$ & -- & -- & \cmark & -- & \cmark\\
       HC$_3$N & -- & -- & \cmark & --& \cmark\\
       H$_2$ & \cmark &  \cmark &  \cmark &  \cmark &  \cmark\\
       \hline
       References & This work & 1 & 2 & 3,4 & 4,5 \\
       \hline
    \end{tabular}
    \tablefoot{
    $^{a}$Values based on stellar and planetary scaling from \cite{betti_near-infrared_2022}. 
    $^{b}$Marginal or weak detections.\\
    References: 
    (1) \cite{long_first_2025},
    (2) \cite{perotti_minds_2025},
    (3) \cite{pascucci_role_2023} for the typical values,
    (4) \citet{grant_transition_2025, arabhavi_minds2_2025} for the molecules detected in averaged MIR spectra,
    (5) Typical VLMS and BD properties comes from the sample describe in \cite{arabhavi_minds2_2025},
    and references therein.
    These targets are selected for their similar properties, such as accretion rate, detected molecules, and age (e.g., J0446B), or for their proximity in mass (e.g., J0438), along with averaged values of T Tauri stars and VLMS observed at MIR \citep{grant_transition_2025, arabhavi_minds2_2025}.}
    \label{tab:comparison_targets}
\end{table*}


\subsection{Evidence of circumplanetary outflows}
\label{sec:wind}
The pure rotational lines of molecular H$_2$ has been detected in many VLMS disks \citep{arabhavi_minds2_2025}.
The detection of H$_2$ emission in T Tauri stars or VLMS can originate from various components, including the surface layers of the disk or from disk-driven winds and outflows \citep{beck_spatially_2008, pascucci_jwstnirspec_2024}.
In the case of Delorme\,1\,AB\,b, the molecular emission appears spatially extended, similar to what is observed 
from class II disks around T Tauri stars \citep{delabrosse_jwst_2024, bajaj_jwst_2024, arulanantham_jwst_2024, schwarz_minds_2025, anderson_gone_2024, pascucci_jwstnirspec_2024}
This spatially resolved H$_2$ emission extends out to approximately 40~au, which is roughly twice the estimated Hill radius. 
This indicates that the outer extent of the emission lies beyond the gravitational sphere of influence of the planet, consistent with outflowing gas.
The spatial resolution of MIRI, combined with the faintness of these lines, limits the ability to conduct a detailed characterization of the wind launching radius or velocity that would provide insight on its origin.

However, this raises the question of how the outflow is launched and how it affects the CPD.
Photoevaporative and MHD disk winds are typically traced using emission lines that probe warm to hot gas, originating from disk layers spanning radii of $\sim$1 to 100 au \citep{ferreira_magnetically-driven_1997, ercolano_modelling_2022,alexander_dispersal_2014,lesur_hydro_2022}
(see also \citealt{pascucci_role_2023} for a recent review).
These winds are launched from disk surfaces, which are irradiated by FUV or X-rays, and generally have velocities ranging from 1 to 5 km/s, with speeds reaching up to a few tens of kilometers per second in cases where the launching region is warmer or closer to the star \citep[see][and references therein]{pascucci_role_2023}.

Assuming that all H$_2$ emission (both resolved and unresolved) is coming from a wind with a uniform mass distribution along the outflow, considering a radial or collimated flow over a distance $r$, we estimate the mass-loss rate as
\begin{equation}
\label{eq:mass_loss_rate}
    \dot{M} = \frac{M_{\text{H}_2} \times v}{r}.
\end{equation}

This equation assumes a steady-state flow, with emission supposed constant along the flow.
Given the relatively low $S/N$ of the observed flow, we carefully selected the smallest aperture that captures the majority of the H$_2$ flux; and the main uncertainty will be about the distribution of the intensity of H$_2$.
We adopted a representative constant wind velocity characteristic of disk winds. Observations of H$_2$ winds in T~Tauri stars typically show velocities of $v = 4$~km\,s$^{-1}$ \citep{gangi_giarps_2020}. 
This CPD is likely a smaller version of T Tauri disks: its keplerian velocity is likley slower at a given distance, but winds are likely launched closer in; so we kept this velocity value to estimate the mass-loss rate.
The radius \texttt{r} corresponds to the inferred emitting size, i.e. 40\,au.
The approximate mass-loss rate measured is $2 \times 10^{-10} \, M_{\mathrm{Jup}}\,\mathrm{yr}^{-1}$.
We note that this H$_2$ reservoir traces only the warm gas, so the actual mass-loss rate could be significantly higher. Although the residual maps reveal clear extended emission, as confirmed by aperture photometry with various aperture sizes, the $S/N$ and spatial resolution are insufficient to reliably characterize the spatial distribution of the flux. The measured masses should therefore be considered as an order-of-magnitude estimate.

The approximate mass ejection rate is around 2$ \times 10^{-10}$ M$_\mathrm{Jup}$/yr, resulting in a wind-to-accretion mass ratio of $\dot{M}_\mathrm{winds}/\dot{M}_\mathrm{acc} \sim 0.004 - 0.1$ \citep[using both scaling methods for accretion values,][]{betti_near-infrared_2022}. 
When the accretion rate is estimated using a “stellar” scaling \citep{betti_near-infrared_2022}, the measured ratio for Delorme\,1\,AB\,b aligns with the values expected for typical Class II disks, despite the system’s older age. 
Indeed, typical Class II disks (ages ranging from 0.3 to 10 Myr) exhibit a wind-to-accretion ratio of $\dot{M}_\mathrm{winds}/\dot{M}_\mathrm{acc} \sim 0.1 - 1$ \citep{pascucci_role_2023}. 
In contrast, when using a “planetary” scaling \citep{betti_near-infrared_2022} to estimate the accretion rate, the wind-to-accretion ratio is found to be one order of magnitude lower.
The measured $\dot{M}_\mathrm{winds}$ for the warm H$_2$ gas in this CPD is slightly lower than typical values for T Tauri stars, which range from $10^{-11}$ to $10^{-8}~M_\odot~\mathrm{yr}^{-1}$ \citep{pascucci_role_2023}, 
for example, Tau\,042021 with $\dot{M}_\mathrm{winds} \sim 6.5 \times 10^{-9}~M_\odot~\mathrm{yr}^{-1}$ \citep{arulanantham_jwst_2024}.

We report the non-detection of any MIR H{\sc i} lines that are associated with accretion in higher mass objects \citep{rigliaco_probing_2015,tofflemire_coordinated_2025}.
A range of additional emission lines were also investigated, including: [Fe\,{\sc ii}] (5.34\,$\mu$m), [Ni\,{\sc ii}] (6.636\,$\mu$m), [Ar\,{\sc ii}] (6.985\,$\mu$m), [Ne\,{\sc ii}] (12.814\,$\mu$m), and [Fe\,{\sc ii}] (17.936\,$\mu$m).
However, none of these lines are detected — neither in the original spectra extracted using various apertures size, nor in the residual spectra obtained after subtracting the best-fit CPD model.
We also searched for spatially extended emission from a variety of molecular species, including CH$_4$ (7.65\,$\mu$m), C$_2$H$_2$ (13.69$\mu$m), $^{13}$CCH$_2$ (13.73\,$\mu$m), C$_2$H$_4$ (10.53\,$\mu$m), C$_2$C$_6$ (12.17\,$\mu$m), C$_3$H$_4$ (15.8\,$\mu$m), C$_4$H$_2$ (15.92\,$\mu$m), C$_6$H$_6$ (14.85\,$\mu$m), and CO$_2$ (15.0\,$\mu$m), as well as excited transitions of CO$_2$ at 13.88\,$\mu$m and 16.18\,$\mu$m, $^{13}$CO$_2$ at 15.41\,$\mu$m and its excited line at 16.2\,$\mu$m, HCN at 14.0\,$\mu$m and its excited line at 14.3\,$\mu$m, HC$_3$N at 15.08\,$\mu$m, and CH$_3$ at 16.48\,$\mu$m.
None of these species are detected in extended emission in these data.

The [Ne\textsc{ii}] line at 12.8\,$\mu$m is a common tracer of photoevaporative winds but can also originate from jets. 
It arises from gas ionized by EUV or X-ray radiation with evidence of a transition from high-velocity, jet-like emission in strong accretors to slower, wind-like emission as accretion decreases \citep[e.g.,][]{pascucci_evolution_2020}.

\subsection{A long-lived primordial disk hypothesis}
The detection of [Ne{\sc ii}] and [Ar{\sc ii}] emission in J0446\,B has been interpreted as evidence of a long-lived primordial disk that has retained its primordial gas content \citep{long_first_2025}.
Interestingly, despite a comparable mass accretion rate ($2.5 \times 10^{-11}$\,M$_\odot$\,yr$^{-1}$), Delorme\,1\,AB\,b shows no such ionized gas tracers, suggesting a markedly different ionization environment. 
This difference is consistent with Delorme\,1\,AB\,b being $\sim$13 times less massive and $\sim$50 times less luminous than J0446\,B. 
The intrinsic faintness of Delorme\,1\,AB\,b compared to other very low-mass stars and T Tauri systems observed with MIRI/MRS complicates direct comparisons.
The disk is likely heated primarily by optical and infrared photons, with limited UV or X-ray irradiation. 
While its properties may still be consistent with a slowly evolving primordial disk — expected around low-mass objects — it is unlikely to be sustained by the planet’s irradiation alone.

\section{Conclusion and perspective}
\label{sec:conclusion}
We confirm the presence of circumplanetary material around the young accreting planet Delorme\,1\,AB\,b 
and unveil for the first time the molecular gas in emission in a CPD thanks to the unique combination of sensitivity and spectral resolution of MIRI/MRS.
These observations also provide hints of outflowing material around an exoplanet.
Our conclusion are as follows:
\begin{itemize}
    \item The atmospheric fitting using visible to MIR data allows us to refine the atmospheric properties of this exoplanet. 
    The best-fit models, derived using four different atmospheric models, provide the following parameters:
    $T_{\text{eff}} = 1725 \pm 134$ K, $R = 1.9 \pm 0.1 \, R_{\text{Jup}}$, a rather super-solar metallicity and C/O ratio
    (although these values are model dependent).
    Molecular absorption of H$_2$O and CO are also detected in the atmosphere of the planet.
    \item The circumplanetary material is modeled as a blackbody with $T_{\text{bb}} = 295 \pm 27 $ K and effective area radius $R = 18.8 \pm 2.7 \, R_{\text{Jup}}$.
    This traces cold dust, and can be interpreted with the presence of a dust cavity that could extend up to 32.6\,R$_\mathrm{Jup}$.
    
    \item 
    We present the first medium-resolution spectrum of a CPD, revealing a carbon-rich molecular gas composition.
    We detect strong HCN emission, characterized by a temperature of 475\,K and a column density of 2.15\,$\times$10$^{17}$ cm$^{-2}$.
    Additionally, we identify C$_2$H$_2$ with both optically thick and thinner components. While the properties of the optically thick component remain unconstrained, we measure a temperature of 650\,K and a column density of 1.47\,$\times$10$^{17}$ cm$^{-2}$ for the thinner component.
    These column densities are approximately one order of magnitude lower than those observed in the VLMS J0446B at a similar age. 
    We tentatively detect the isotopologue $^{13}$CCH$_2$.

    \item 
    Two H$2$ emission lines are detected in extended emission, reaching distances up to approximately twice the Hill radius. 
    If this emission traces a wind dispersing the disk, we estimate a mass-loss rate of $2 \times 10^{-10}$\,M$_{\mathrm{Jup}}.\mathrm{yr}^{-1}$.
    Assuming all the observed H$_2$ originates from the wind, this implies a wind-to-accretion mass ratio of $\dot{M}_{\mathrm{wind}}/\dot{M}_{\mathrm{acc}}$ $\sim$ 0.004 - 0.1, depending on the adopted accretion rate.
    
\end{itemize}
These results present a challenge to current planet formation theories, as they imply that significant amounts of gas remain present in the system even after $\sim$30\,Myr; well beyond the typical gas dispersal timescales for circumstellar disks.
The persistence of this gas-rich environment supports the existence of a circumplanetary disk at an advanced age.
Moreover, our analysis reveals that this potentially moon-forming disk is carbon-rich, which could have important implications for satellite formation. 
A high carbon content may influence the composition and chemical diversity of any forming moons, as indicated for the solar system \citep[e.g. for Jupiter's protolunar disk,][]{canup_formation_2002}.

\begin{acknowledgements}
    We thank the anonymous reviewer for providing a positive and constructive report.
    This paper is based on observations made with the NASA/ESA/CSA James Webb Space Telescope and   reports work carried out in the context of the \emph{JWST} Telescope Scientist Team (PI M.~Mountain).
    The data were obtained from the Mikulski Archive for Space Telescopes at the Space Telescope Science Institute, which is operated by the Association of Universities for Research in Astronomy, Inc., under NASA contract NAS 5-03127 for JWST. 
    This data can be accessed via DOI \href{https://doi.org/10.17909/v03e-qk60}{doi:10.17909/v03e-qk60}.
    These observations are associated with program GTO 2278 (PI: M.~Perrin).  
    Support for program 2278 was provided by NASA through a grant from the Space Telescope Science Institute, which is operated by the Association of Universities for Research in Astronomy, Inc., under NASA contract NAS 5-03127.
\end{acknowledgements}

\bibliographystyle{aa}
\bibliography{aa56792-25}

\begin{appendix}

\section{Atmospheric fitting with various atmosphere models}
\label{sec:appendix_atmos}
This section provides additional figures supporting Section~\ref{sec:atm_charact}. 
Table~\ref{tab:param_models_atm} shows the bounds for each parameter for all model grids.
For all model fits, effective temperatures were restricted to 1500--2000K and object radii allowed to range from 0.5--10\,$R_\mathrm{Jup}$.
\begin{table*}[t]
    \centering
    \caption{Parameters of the atmospheric grids.}
    \begin{tabular}{c|c c c c}
        \hline
        \hline
        Parameters & \texttt{Exo-REM} &  \texttt{ATMO} &  \texttt{Sonora diamondback} &  \texttt{BT-Settl CIFIST}\\
        \hline
        \hline
        Temperature (K)  & 400 -- 2000  & 800 -- 3000 & 900 --  2400 & 1200--7000\\
                         & Step: 50    & Step: 100 & Step: 100 & Step: 100\\
        \hline                 
        $\mathrm{Log} (g)$ &  3.0 -- 5.0  & 2.5 -- 5.5 & 3.5 -- 5.5 & 3.0--5.5\\
                         & Step: 0.5  & Step: 0.5 & Step: 0.5 & Step: 0.5\\ 
        \hline
        C/O              &  0.1 -- 0.8  &  0.3, 0.55, 0.7 & -- & -- \\
                         & Step: 0.05 & -- & -- & -- \\
        \hline
        Metallicity   & -0.5 -- 1 & -0.6 -- 0.6 & -0.5 -- 0.5 & -- \\
              & Step: 0.5 & Step: 0.3 & Step: 0.5 & -- \\
       \hline
        $\gamma$         &    --      & 1.01, 1.03, 1.05 & -- & --\\
        \hline
        f$_{sed}$ & -- & -- & [1, 2, 3, 4, 8, nc] & -- \\
        \hline
        \end{tabular}
        \tablefoot{$\gamma$ is specific to the \texttt{ATMO} grid and corresponds to the adiabatic index. f$_{sed}$ is used to parametrize the clouds in the \texttt{Sonora} grid (with nc being non-cloud).}
    \label{tab:param_models_atm}
\end{table*}

\begin{figure*}[ht]
    \centering
    \includegraphics[width=1\linewidth]{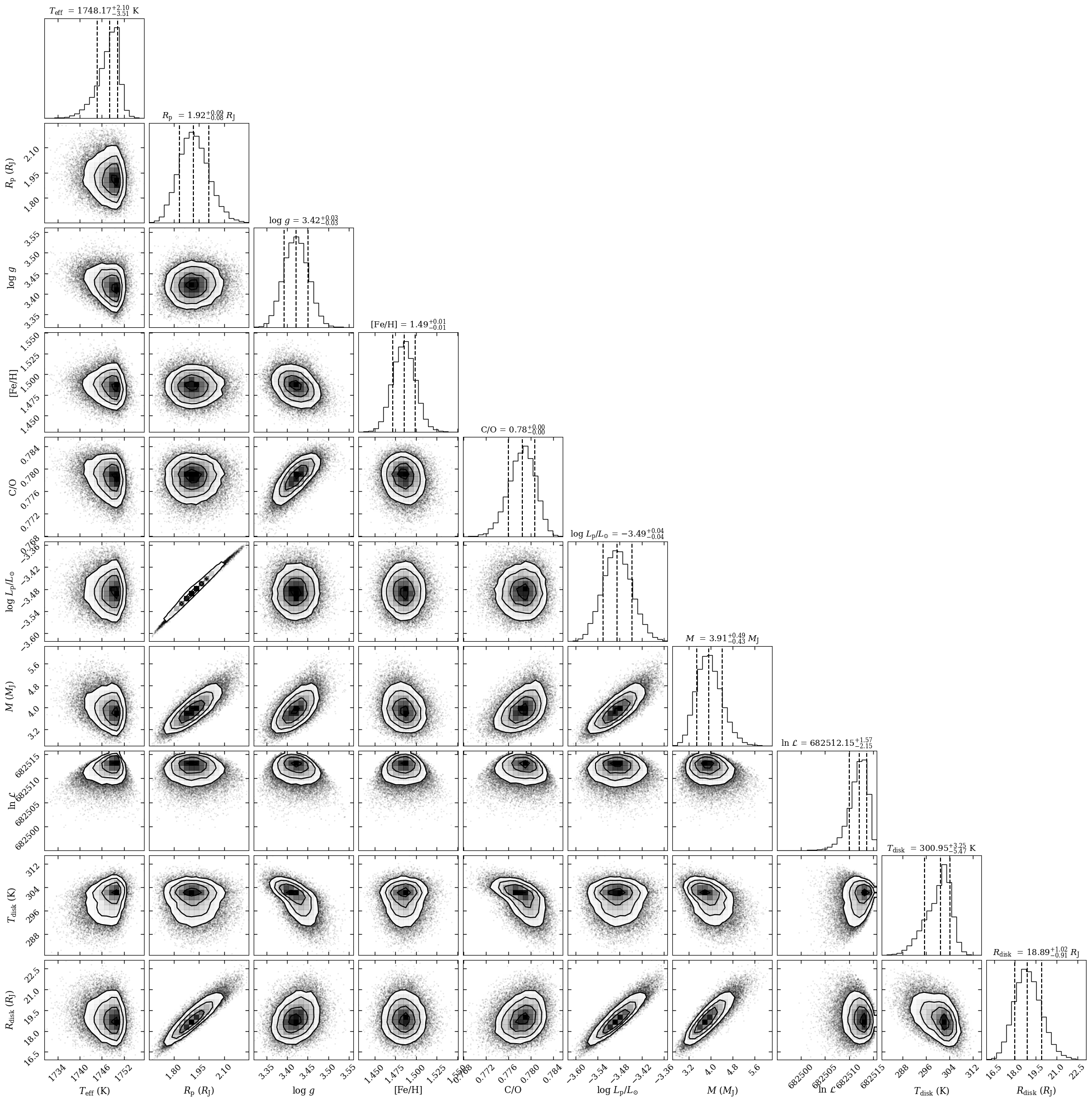}
    \caption{Corner plot corresponding to the \texttt{Exo-REM} model shown in  Figure~\ref{fig:best_fits_exorem}.}
    \label{fig:cp_exorem}
\end{figure*}

\begin{figure*}[ht]
    \centering
    \includegraphics[width=1\linewidth]{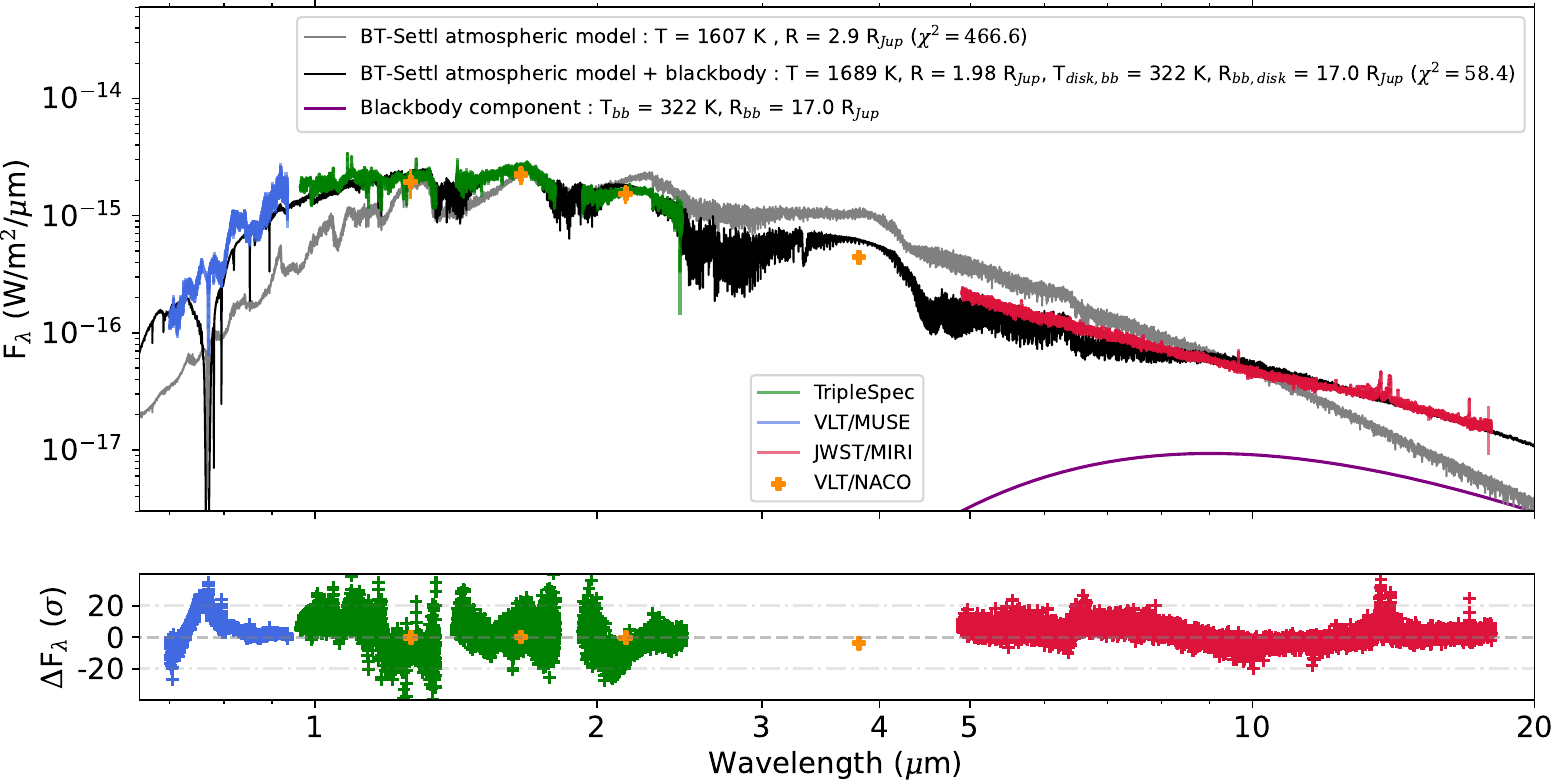}
    \caption{Spectrum and photometric data of Delorme\,1\,AB\,b from the UV to the MIR, shown together with the best-fit \texttt{BT-Settl} model (as in Figure~\ref{fig:best_fits_exorem}).}
    \label{fig:best_fits_btsettl}
\end{figure*}

\begin{figure*}[ht]
    \centering
    \includegraphics[width=1\linewidth]{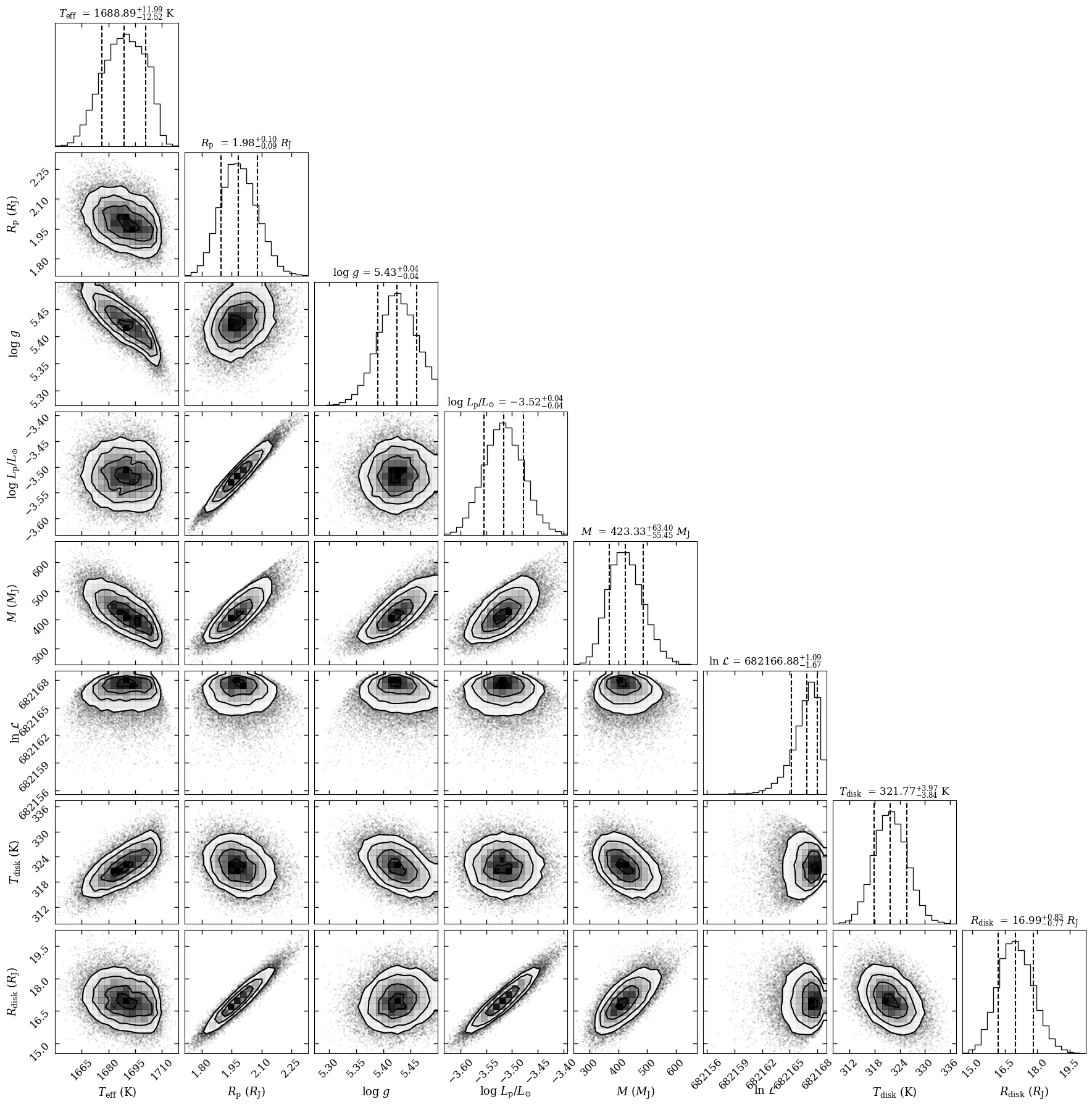}
    \caption{Corner plot corresponding to the best-fit \texttt{BT-Settl} model shown in  Figure~\ref{fig:best_fits_btsettl}.}
    \label{fig:cp_btsettl}
\end{figure*}

\begin{figure*}[ht]
    \centering
    \includegraphics[width=1\linewidth]{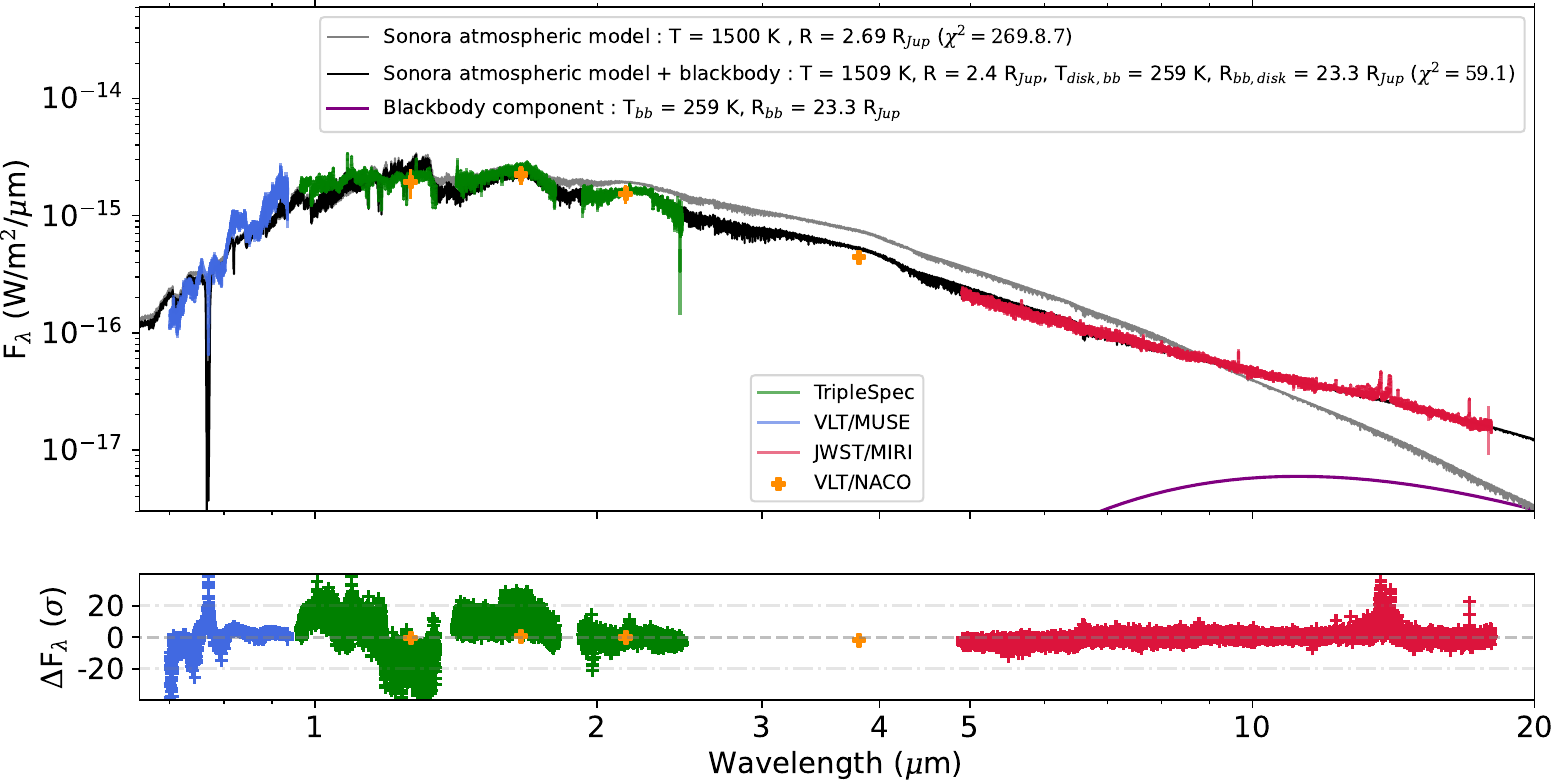}
    \caption{Spectrum and photometric data of Delorme\,1\,AB\,b from the UV to the MIR, shown together with the best-fit \texttt{Sonora} model (as in Figure~\ref{fig:best_fits_exorem}).}
    \label{fig:best_fits_sonora}
\end{figure*}

\begin{figure*}[ht]
    \centering
    \includegraphics[width=1\linewidth]{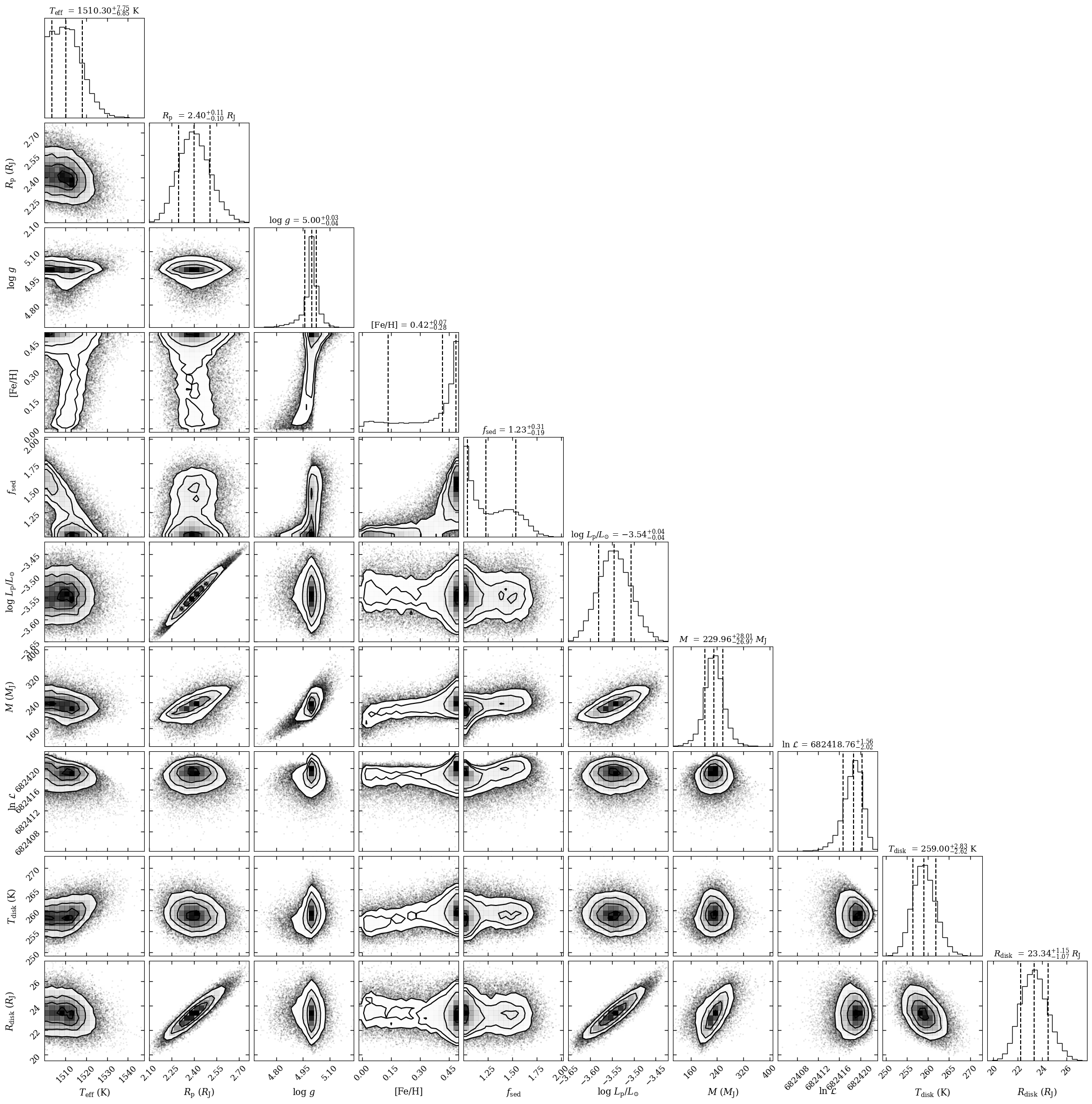}
    \caption{Corner plot corresponding to the best-fit \texttt{Sonora} shown in  Figure~\ref{fig:best_fits_sonora}.}
    \label{fig:cp_sonora}
\end{figure*}

\begin{figure*}[ht]
    \centering
    \includegraphics[width=1\linewidth]{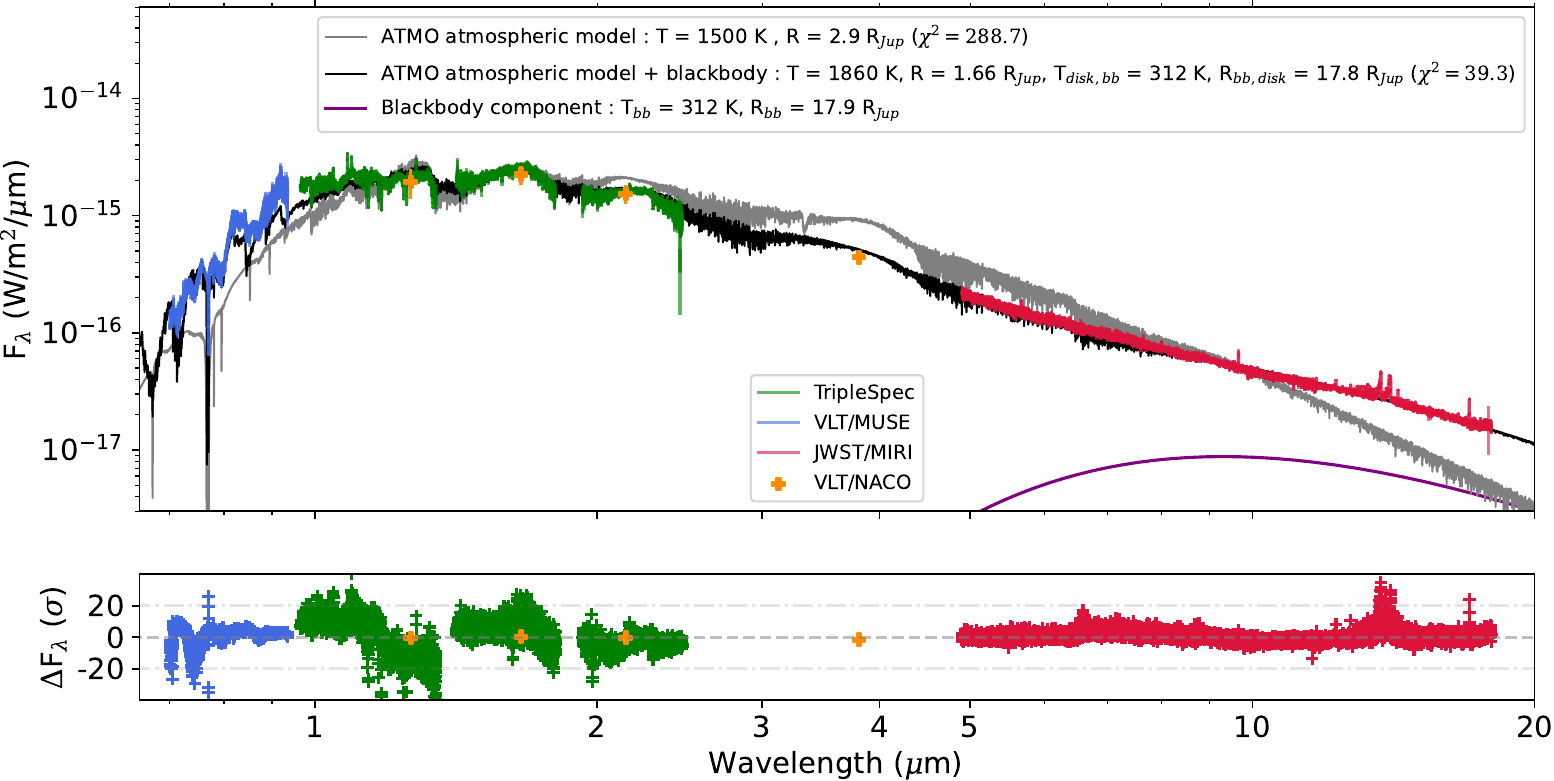}
    \caption{Spectrum and photometric data of Delorme\,1\,AB\,b from the UV to the MIR, shown together with the best-fit \texttt{ATMO} model (as in Figure~\ref{fig:best_fits_exorem}).}
    \label{fig:best_fits_atmo}
\end{figure*}

\begin{figure*}[ht]
    \centering
    \includegraphics[width=1\linewidth]{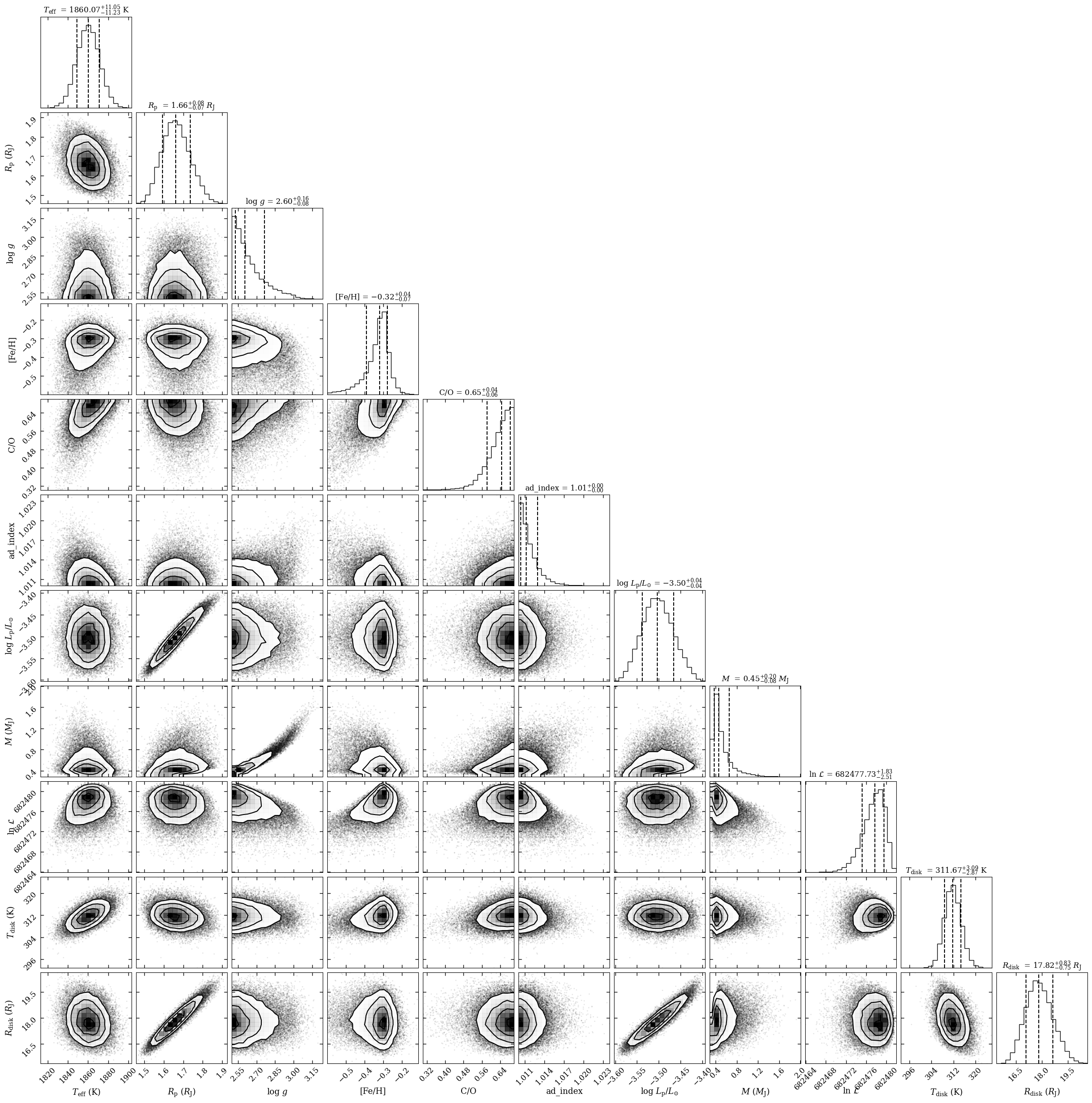}
    \caption{Corner plot corresponding to the best fit \texttt{ATMO} shown in  Figure~\ref{fig:best_fits_atmo}.}
    \label{fig:cp_atmo}
\end{figure*}

The best-fit models and their residuals are shown alongside the data from UV to MIR, incorporating the blackbody component (Figures~\ref{fig:best_fits_btsettl}, \ref{fig:best_fits_sonora}, and \ref{fig:best_fits_atmo}). 
The corresponding corner plots are shown in Figures~\ref{fig:cp_exorem}, \ref{fig:cp_btsettl}, \ref{fig:cp_sonora}, and \ref{fig:cp_atmo}




\begin{figure*}[ht]
    \centering
    \includegraphics[width=1\linewidth]{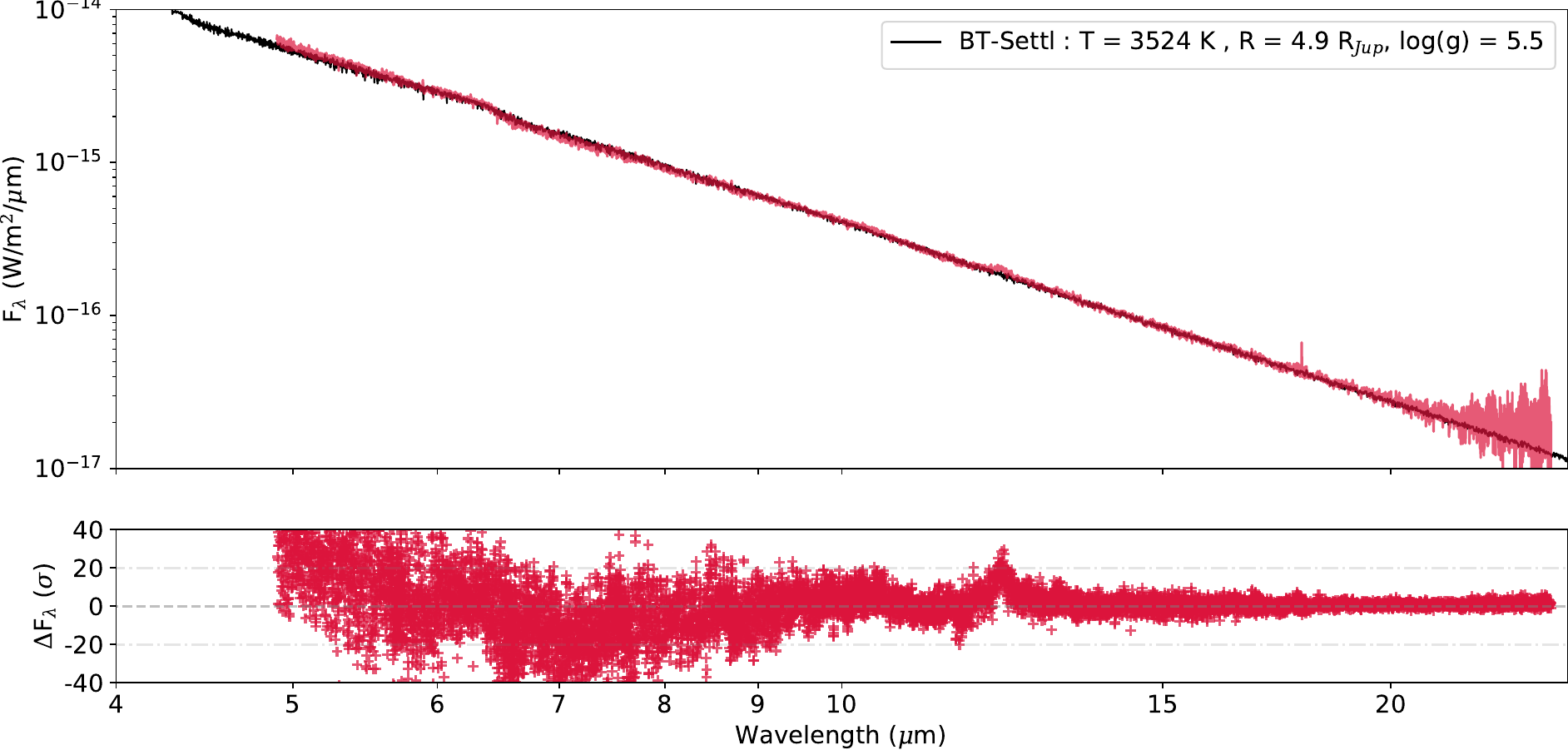}
    \caption{Spectrum and photometric data of the host binary star Delorme\,1\,AB shown together with the best-fit \texttt{BT-Settl} model.}
    \label{fig:best_fits_host}
\end{figure*}

\begin{figure*}[ht]
    \centering
    \includegraphics[width=0.5\linewidth]{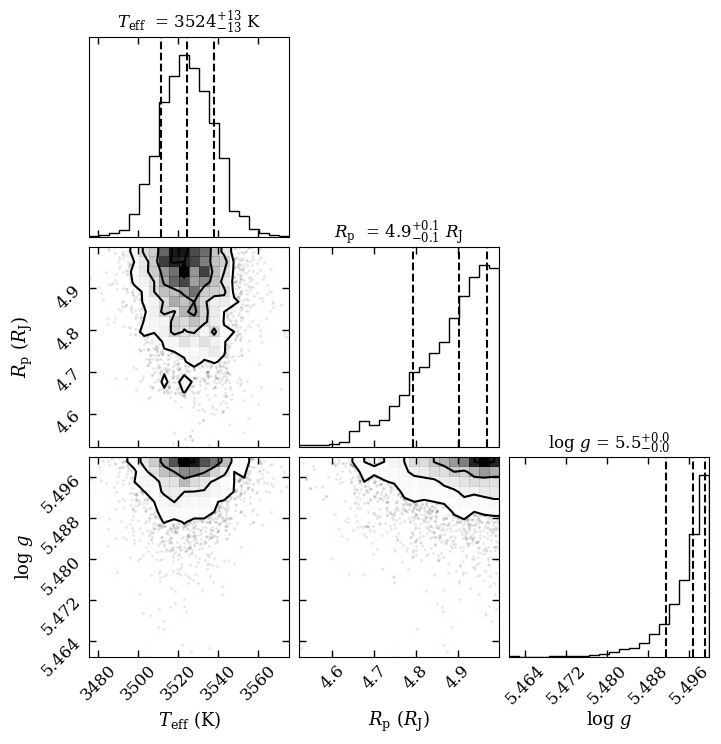}
    \caption{Corner plot corresponding to Fig. \ref{fig:best_fits_host}.}
    \label{fig:best_fits_host_cp}
\end{figure*}

Figure \ref{fig:best_fits_host} shows the best-fit models overlaid on the MIRI/MRS data for the binary host star, and Fig.\,\ref{fig:best_fits_host_cp} presents the corresponding corner plot.

\begin{figure*}[ht]
    \centering
    \includegraphics[width=0.75\linewidth]{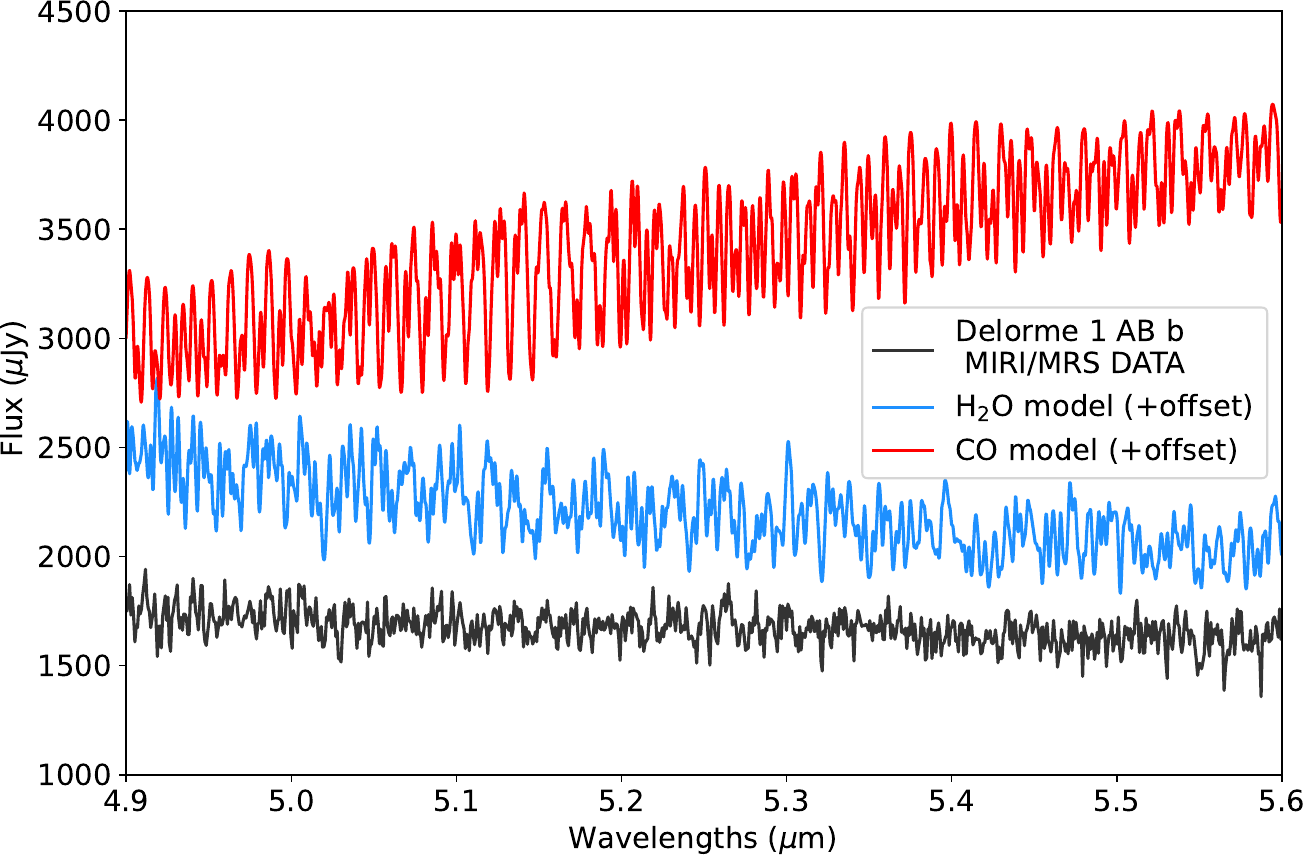}
    \caption{Molecular template spectra of H$_2$O and CO. The spectrum of Delorme\,1\,AB\,b is shown in black for comparison.}\label{fig:Spectrum_H2O_CO}
\end{figure*}

Finally, Figure \ref{fig:Spectrum_H2O_CO} presents the spectrum of Delorme\,1\,AB\,b alongside the model spectra for CO and H$_2$O. 
These molecular templates are computed using the pressure–temperature profile corresponding to the best-fit parameters of the \texttt{Exo-REM} atmospheric model.
They are calculated under chemical equilibrium, based on previously derived abundance profiles.
The radiative transfer is then recomputed with \texttt{Exo-REM}, including only the molecular species of interest, while retaining collision-induced absorption (H\textsubscript{2}–H\textsubscript{2}, H\textsubscript{2}–He, H\textsubscript{2}O–H\textsubscript{2}O).

\section{Correlation analysis in 2D: Molecular maps.}
\label{sec:molecular_mapping}
The ``molecular mapping'' is applied exactly as described in \cite{malin_simulated_2023}, with adapted $S/N$ measurements focused on the companion, i.e., by masking the host component in the noise calculation.
\subsection{Atmospheric molecular templates}

\begin{figure*}
    \centering
    \includegraphics[width=1\linewidth]{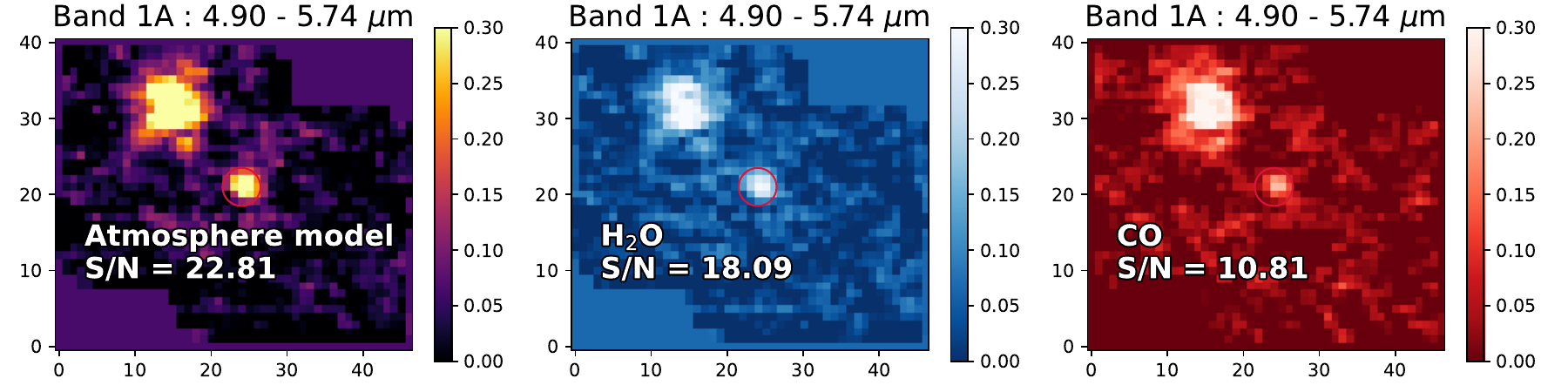}
    \caption{Correlation maps. The first panel (left) shows the cross correlation values for each pixel between the spaxel (pixel spectrum) and the \texttt{Exo-REM} best-fit model. The middle panel shows the correlation values with a template model spectrum of H$_2$O only, while the right panel uses a CO model template.}
    \label{fig:molecular_map_atm}
\end{figure*}

First, the correlation of the spectral  cube with atmospheric molecular templates (using \texttt{Exo-REM} models at 1750\,K), we achieve strong detections of CO and H$_2$O, as expected.
These molecules are detected in both the companion and the M-star binary host.
The correlation maps are displayed in Figure~\ref{fig:molecular_map_atm}.

\subsection{Hydrocarbons emission}
\begin{figure*}
    \centering
    \includegraphics[width=1\linewidth]{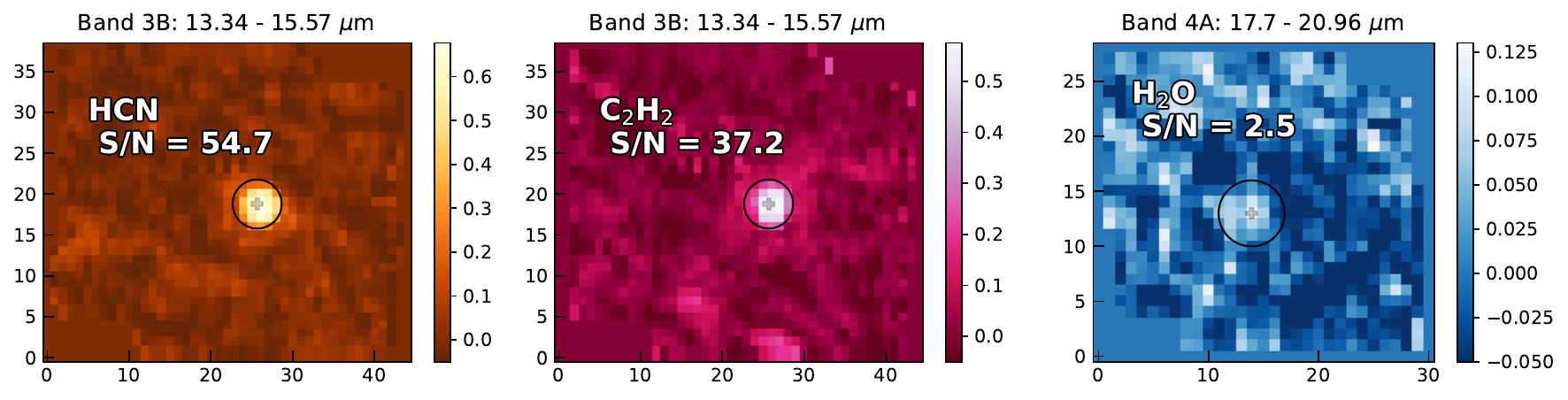}
    \caption{Correlation maps. The first two panels show the cross correlation values for each pixel between the spaxel and the hydrocarbon templates: HCN (left) and C$_2$H$_2$ (middle). The right panel shows the correlation map with a template of H$_2$O in emission.
    The $S/N$ values are displayed for the companion.}
    \label{fig:molec_map_cpd}
\end{figure*}

\begin{figure}
    \centering
    \includegraphics[width=1\linewidth]{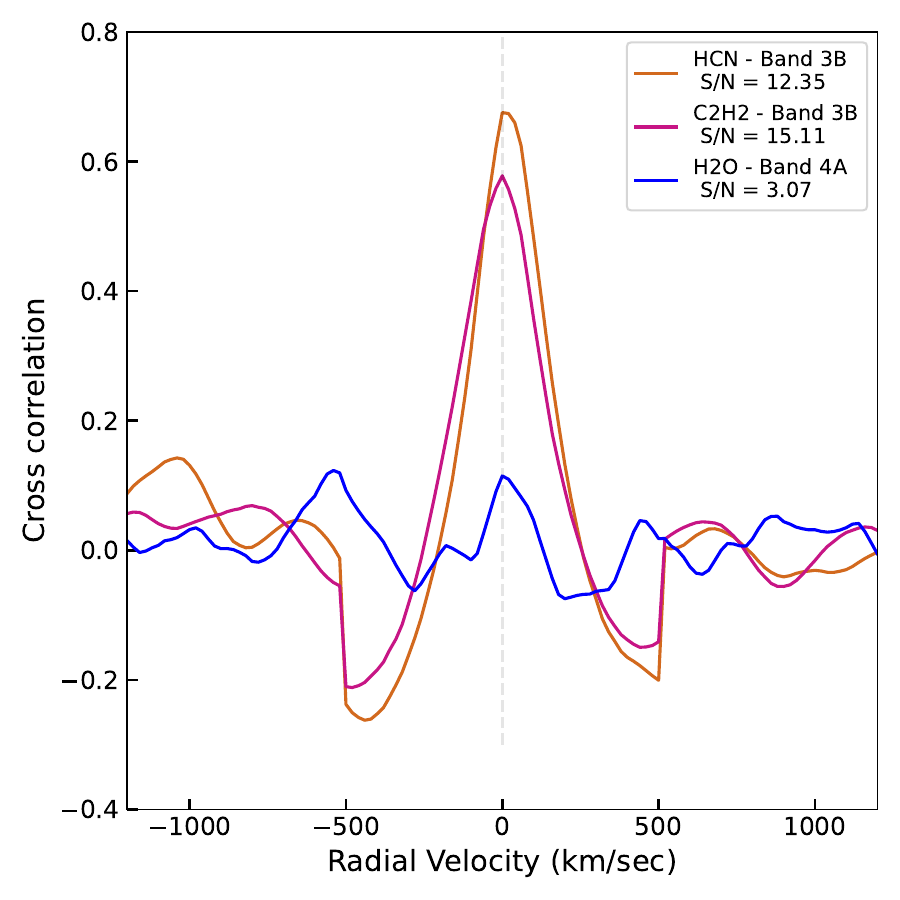}
    \caption{Cross correlation with the "slab" molecular emission templates (Section~\ref{sec:cpd_molec}). Strong detections of HCN and C$_2$H$_2$ are observed in band 3B, along with a weak detection of H$_2$O.}
    \label{fig:CCF_CPD}
\end{figure}

Similarly, we apply the ``molecular mapping'' method using model spectra of hydrocarbons: we correlate the data cubes with each hydrocarbon mode.
The resulting 2D correlation maps are shown in Figure~\ref{fig:molec_map_cpd} for the detected species. 
For comparison, we include the 1D CCF with the companion's spectrum, as shown in Figure~\ref{fig:CCF_CPD}.
This correlation reveals a very low-$S/N$ detection of H$_2$O in emission in band 4A (wavelengths from $\sim$ 17.7–21 $\mu$m), although it is too weak to be confirmed.

\section{Comparison with evolution models}
\begin{figure}[ht]
    \centering
    \includegraphics[width=1\linewidth]{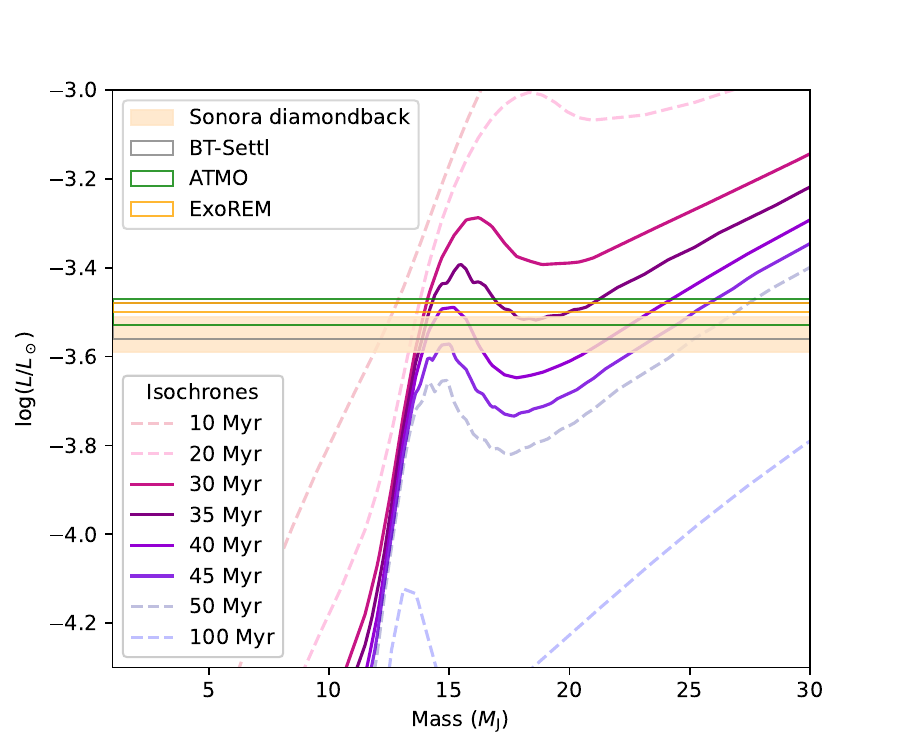}
    \includegraphics[width=1\linewidth]{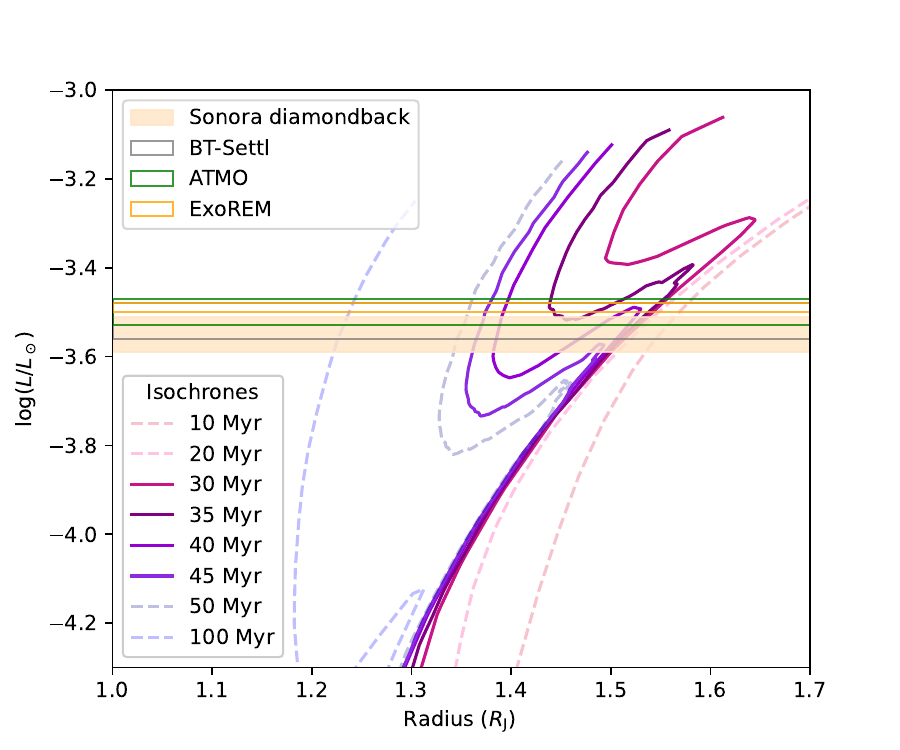}
    \caption{Measured luminosity (with the four different atmospheric models) as a function of the mass (in M$_\mathrm{Jup}$), and radius (in R$_\mathrm{Jup}$).
    Isochrones come from the \texttt{sonora} models. Results are similar with \texttt{ATMO} evolutionary models.}
    \label{fig:evolution}
\end{figure}

Figure~\ref{fig:evolution} presents the best-fit luminosity values for Delorme\,1\,AB\,b, which can be used to estimate its mass and radius by comparing with evolutionary models (here using \texttt{Sonora} models). 
These values can subsequently serve as priors before performing atmospheric fits, as described in Section~\ref{sec:atm_charact}. Table~\ref{tab:best_fits_atm_CPD_priors} shows the atmospheric properties derived when including priors from evolutionary models.

\begin{table*}[h!]
    \caption{Summary of the best-fit parameters for the planetary mass companion measured with different atmospheric grids and including priors on the mass and radius from evolutionary models, before running the fits.}
    \centering
    \begin{tabular}{c|ccccccc:cc:c}
        \hline
        \hline
         Model & T$_{eff}$ & R (R$_\mathrm{Jup}$) & log(g) & [Fe/H] & C/O & f$_{sed}$ & $\gamma$ &
         T$_{bb}$ & R$_{bb}$ (R$_\mathrm{Jup}$) & $\chi^2$\\ 
         \hline
         \hline
         
         \texttt{Exo-REM} 
         & 1786$^{+2}_{-2}$ & 
         1.97$^{+0.03}_{-0.03}$ & 
         3.92$^{+0.01}_{-0.01}$ &
         1.48$^{+0.01}_{-0.01}$ &
         0.80$^{+0.01}_{-0.01}$ &
         -- &
         -- &
         -- & 
         -- &
         298.3\\

        \texttt{Exo-REM} + BB
         & 1899$^{+4}_{-5}$ & 
         1.78$^{+0.03}_{-0.03}$ & 
         4.00$^{+0.01}_{-0.01}$ &
         1.47$^{+0.01}_{-0.01}$ &
         0.72$^{+0.01}_{-0.01}$ &
         -- &
         -- &
         314$^{+3}_{-3}$ & 
         18.4$^{+0.4}_{-0.4}$ &
         41.9\\

        \hline
         \texttt{ATMO} 
         & 1560$^{+0.1}_{-0.6}$ & 
         1.91$^{01}_{-0.01}$ & 
         3.83$^{+0.01}_{-0.01}$ &
         -0.57$^{+0.01}_{-0.01}$ &
         0.67$^{+0.01}_{-0.01}$ &
         -- &
         1.03$^{+0.01}_{-0.01}$ &
         -- & 
         -- &
         301.4\\

         \texttt{ATMO} + BB
         & 1832$^{+9}_{-9}$ & 
         1.72$^{+0.04}_{-0.05}$ & 
         4.04$^{+0.03}_{-0.03}$ &
         -0.59$^{+0.02}_{-0.01}$ &
         0.69$^{+0.01}_{-0.01}$ &
         -- &
         1.02$^{+0.01}_{-0.01}$ &
         311$^{+3}_{-3}$ & 
         17.9$^{+0.5}_{-0.5}$ &
         43.9\\

         \hline
         \texttt{Sonora} 
         & 1500$^{+1}_{-1}$ & 
         2.09$^{+0.01}_{-0.01}$ & 
         3.6$^{+0.01}_{-0.01}$ &
         0.40$^{+0.01}_{-0.01}$ &
         -- &
         1.62$^{+0.01}_{-0.01}$ &
         -- &
         -- & 
         -- & 
         495.5\\
        \texttt{Sonora} + BB
         & 1557$^{+2}_{-2}$ & 
         1.78$^{+0.1}_{-0.1}$ & 
         4.07$^{+0.01}_{-0.01}$ &
         0.49$^{+0.01}_{-0.01}$ &
         -- &
         1.03$^{+0.01}_{-0.01}$ &
         -- &
         293$^{+1}_{-1}$ & 
         15.4$^{+0.06}_{-0.06}$ & 
         110.9 \\

        \hline
        \texttt{BT-Settl} 
         & 1609$^{+0.1}_{0.1}$ & 
         1.84$^{+0.01}_{-0.01}$ & 
         3.77$^{+0.01}_{-0.01}$ &
         -- &
         -- &
         -- &
         -- &
         -- & 
         -- &
         437.3\\

        \texttt{BT-Settl} + BB
         & 1742$^{+4}_{-3}$ & 
         1.54$^{+0.03}_{-0.03}$ & 
         4.19$^{+0.02}_{-0.02}$ &
         -- &
         -- &
         -- &
         -- &
         324$^{+3}_{-3}$ & 
         14.4$^{+0.4}_{-0.3}$ &
         148.3\\
        \hline
    \end{tabular}
    \tablefoot{The second row for each model includes a blackbody component.}
    \label{tab:best_fits_atm_CPD_priors}
\end{table*}

\section{Continuum estimation}
\begin{figure*}
    \centering
    \includegraphics[width=18cm]{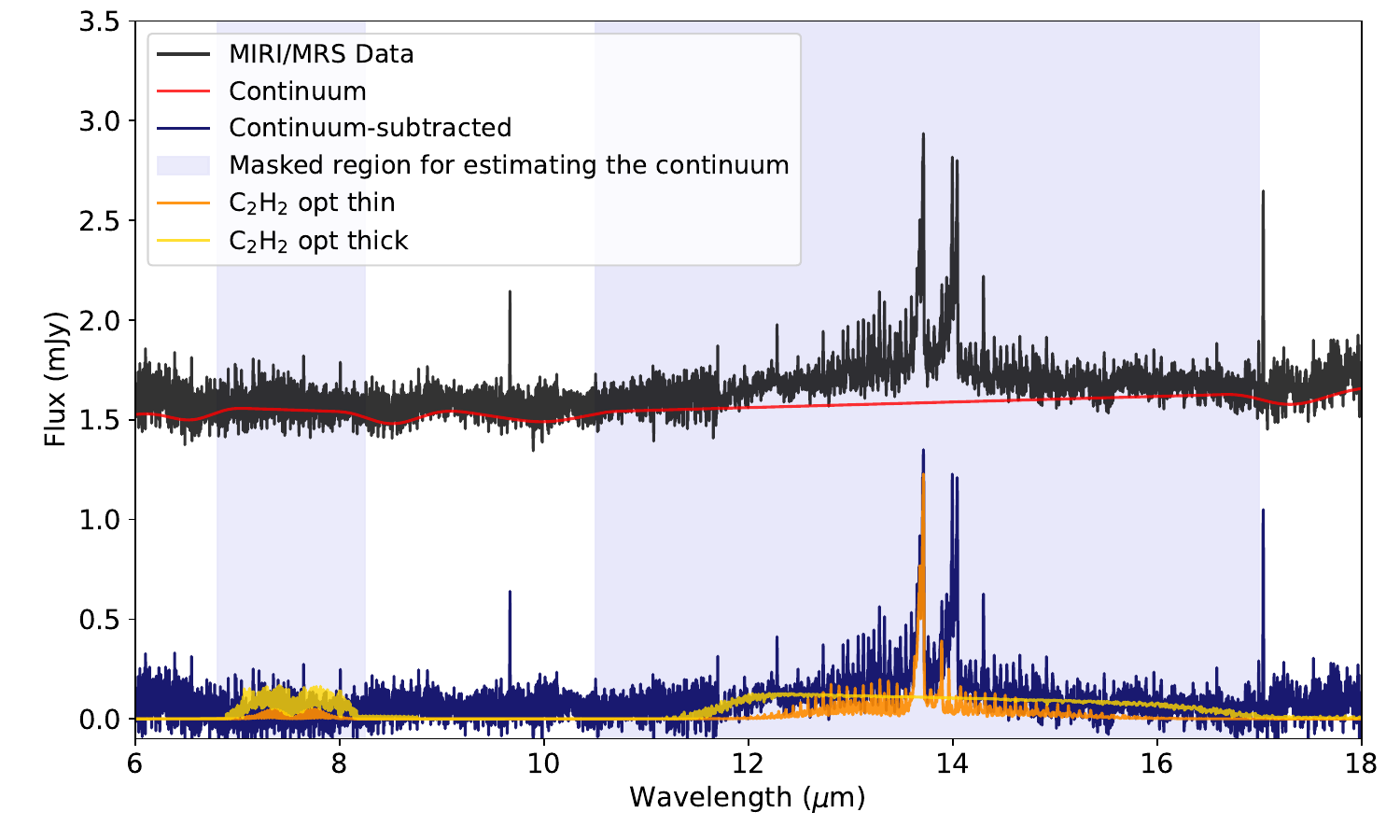}
    \caption{MIRI/MRS spectrum of Delorme 1 AB b in black, with the measured continuum overplotted in red.
    The lower panel shows the continuum-subtracted spectrum together with both C$_2$H$_2$ components.
    Inclusion of both components is required to reproduce the observed C$_2$H$_2$ feature.}
\label{fig:Delorme1ABb_CPD_C2H2_continuum}
\end{figure*}

Figure \ref{fig:Delorme1ABb_CPD_C2H2_continuum} illustrates the challenges of continuum subtraction in the presence of a pseudo-continuum due to the optically thick C$_2$H$_2$ component.
While this effect is less pronounced than in J160532 \citep[Fig. 1][]{tabone_rich_2023}, it can still influence the reliability of the continuum estimation across the 12–16 $\mu$m range.

\section{$\chi^{2}$ Maps for each molecule in the CPD}
\label{sec:chi2maps}
The noise level to determine the 1-, 2-, and 3$\sigma$ contours in the $\chi^2$ plots is determined by taking the standard deviation of the spectrum from 16.15 to 16.25 $\mu$m, resulting in a noise level of 0.041 mJy.
For all figures: red, orange, and yellow contours denote the 1, 2, and 3 $\sigma$ levels.
The white lines denote a equivalent emitting radii in au as labeled.

\begin{figure*}
    \centering
    \includegraphics[width=18cm]{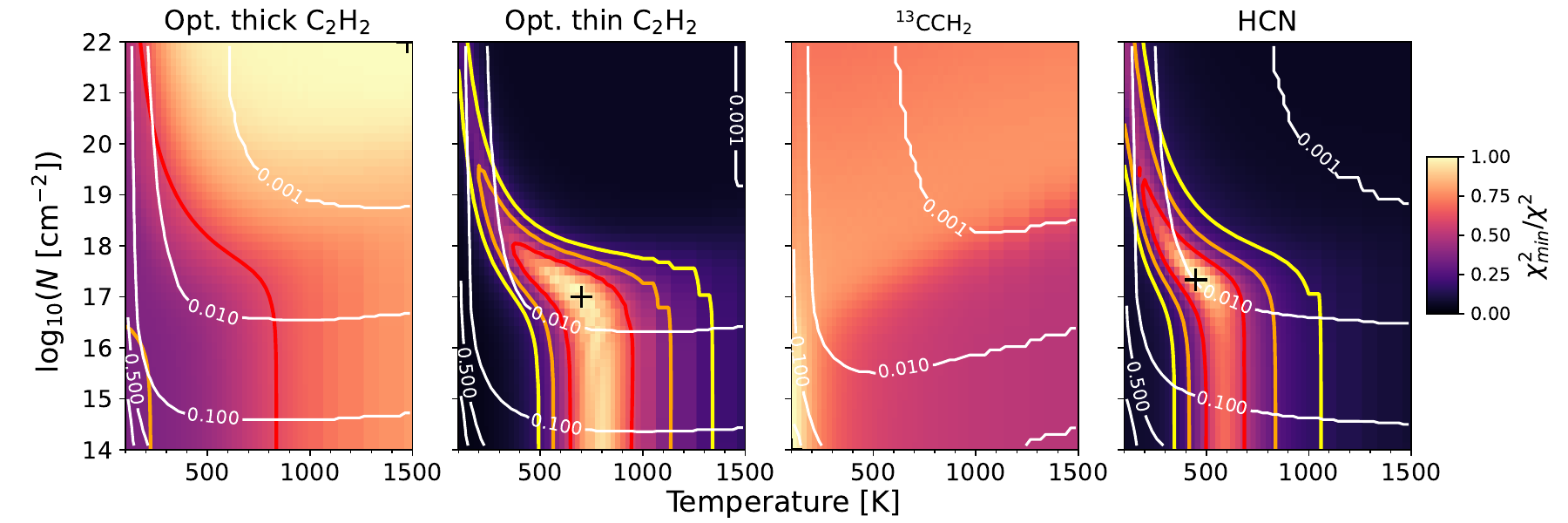}
    \caption{Constraints obtained from $\chi^{2}$ fits for the optically thick component of C$_2$H$_2$, the optically thin component of C$_2$H$_2$, $^{13}$CCH$_2$ and HCN.
    Red, orange, and yellow contours denote the 1, 2, and 3 $\sigma$ levels.
    The white lines denote a equivalent emitting radii with the given values in au.}
    \label{fig:chi2_slabs}
\end{figure*}
\end{appendix}
\end{document}